\documentclass[11pt,a4paper]{article}
\usepackage{jheppub}

\usepackage{graphicx}                     
\usepackage{amsmath,amssymb,amsfonts}
\usepackage{xspace}                       
\usepackage{float}
\usepackage{dcolumn}
\usepackage{fancyhdr}
\usepackage{citesort}
\usepackage{mathtools}
\usepackage{braket}
\usepackage{mathtools, slashed}
\usepackage{dsfont}

\newcommand{\beq}{\begin{equation}}
\newcommand{\eeq}{\end{equation}}
\newcommand{\beqa}{\begin{eqnarray}}
\newcommand{\eeqa}{\end{eqnarray}}
\newcommand{\Order}{\mathcal{O}}

\newcommand{\nn}{\nonumber\\}

\title{Flavor decomposition of the pion-nucleon {\boldmath$\sigma$}-term}

\author[a]{Daniel Severt,}
           
\author[a,b,c]{Ulf-G. Mei{\ss}ner} 

\author[d,c]{and Jambul Gegelia}

\affiliation[a]{Helmholtz-Institut f\"ur Strahlen- und Kernphysik  and Bethe
  Center for Theoretical Physics,\\
           Universit\"at  Bonn,
           D-53115 Bonn, Germany}

\affiliation[b]{Institute for Advanced Simulation (IAS-4) Institut f\"ur
         Kernphysik (IKP-3) and JCHP\\ Forschungszentrum J\"ulich,
         D-52425 J\"ulich, Germany}

\affiliation[c]{Tbilisi State  University,  0186 Tbilisi, Georgia}

\affiliation[d]{ Fakult\"at f\"ur Physik und Astronomie,
  Institut f\"ur Theoretische Physik II\\
  Ruhr-Universit\"at Bochum,
  D-44870 Bochum, Germany}

\emailAdd{severt@hiskp.uni-bonn.de}
\emailAdd{meissner@hiskp.uni-bonn.de}
\emailAdd{jgegelia@hotmail.com}

\abstract{We re-analyze the flavor decomposition of the pion-nucleon $\sigma$-term
in the framework of baryon chiral perturbation to fourth order. We employ a covariant
and the heavy baryon framework including also the low-lying decuplet. Using only
continuum data, we find a small strangeness content of the proton. The uncertainties are,
however, large and might be overcome by dedicated lattice QCD calculations.
}
\preprint{}

\begin{document}
\maketitle

\section{Short introduction}
\label{sec:intro}
\setcounter{footnote}{0}

The pion-nucleon $\sigma$-term $\sigma_{\pi N}$ parameterizes the scalar couplings of the
nucleon to the light up- and down-quarks. It also plays a key role in the search for
physics beyond the Standard Model, such as direct-detection searches for dark matter, see
e.g.~\cite{Bottino:1999ei,Ellis:2008hf,Crivellin:2013ipa}, but also other searches that
are sensitive to the  scalar current coupling to nucleons,
see e.g.~\cite{Crivellin:2014cta,deVries:2016jox}.

Of particular interest is its
flavor decomposition, in which one rewrites the $\sigma$-term in terms of an SU(3)
singlet $\sigma_0$ and the so-called strangeness fraction $y$ as $\sigma_{\pi N} = \sigma_0/(1-y)$.
It is the quantity $\sigma_0$ that will be the central object of this study. In fact,
as will be discussed later, since there is a discrepancy between dispersion theoretical and
lattice QCD determinations of $\sigma_{\pi N}$, it is of interest to analyze  $\sigma_0$ based on
continuum data only. As we will see, the complete one-loop calculations of $\Order(p^4)$
utilizing various formulations of baryon chiral perturbation
theory (also including the decuplet) as done here allows one to pin down  $\sigma_0$ more precisely
than the already available leading one-loop $\Order(p^3)$ calculations. The other novelty of
our calculation is a better estimate of the theoretical uncertainty, not only due to the
parameter variations within a given order but also due to the neglected higher orders.
Not surprisingly, we find that the latter are quite sizeable at third order but much smaller
than the errors within the order for the fourth order calculations.

Needless to say that the formalism developed here can also be applied to analyze the results
of lattice QCD calculations at varying light and strange quark masses. Given the tension
in the value of  $\sigma_{\pi N}$ alluded to before, we refrain from doing that here.
Ultimately, however, we believe that lattice QCD will allow for a more precise determination
of the flavor decomposition of $\sigma_{\pi N}$.

The paper is organized as follows. 
We give the basic definitions concerning the   pion-nucleon $\sigma$-term, the method of calculation
and a brief recapitulation of known results  in Sec.~\ref{sec:def}.
In Sec.~\ref{sec:formalism} we present the chiral Lagrangians necessary for our calculation,
and discuss constraints on various low-energy constants.
Sec.~\ref{SecMassCalc} gives details on the calculations of the baryon masses and the sigma-term
at second, third and fourth order, respectively. The fit procedure and error analysis methods
are discussed in Sec.~\ref{sec:fits}. The results and corresponding discussions are given
in Sec.~\ref{sec:results}. We end with our conclusions in Sec.~\ref{sec:conclusions}.
Various technicalities and formulas are relegated to the appendices.


\section{Sigma term basics}\label{sec:def}

\subsection{Definitions}

The pion-nucleon sigma-term $\sigma_{\pi N}$ is defined as the expectation value of the
light flavor ($u,d$) QCD quark mass term in the nucleon,
\begin{equation}
\sigma_{\pi N} = \frac{\hat{m}}{2 m_N} \bra{N} \bar{u} u + \bar{d} d \ket{N} \; , 
\end{equation}
where $u$ and $d$ are the up- and down-quark fields, respectively, and $\ket{N}$ is a properly
normalized nucleon state,  i.e. $ \braket{N | N} =1 $, with mass $m_N = 938.9\,$MeV.
In what follows,  we restrict ourselves to the isospin limit  $m_u = m_d = \hat{m}$. 
 This approximation is justified, because the masses of the $u$ and $d$ quark are very small compared to 
 $\Lambda_{\mathrm{QCD}} \sim {250}$~MeV and  this assumption simplifies the  calculations.
 There is also  a sigma-term 
 related to the strange quark field
\begin{align}
\sigma_s =  \frac{m_s}{2 m_N} \bra{N} \bar{s} s \ket{N} \; ,
\end{align}
with the strange quark mass $m_s$. One can define  another expression that characterizes
the scalar nucleon  structure, namely the strangeness content of the nucleon $y$. It is defined by
\begin{equation}
y = \frac{2 \bra{N} \bar{s} s \ket{N}}{\bra{N} \bar{u} u + \bar{d} d \ket{N}} 
= \frac{2 \hat{m}}{m_s} \frac{\sigma_s}{\sigma_{\pi N}} \; .
\end{equation} 
In order to calculate the strangeness content, one usually rewrites the $\pi N$ sigma-term
in the following way: 
\begin{align}
\sigma_{\pi N} = \frac{\sigma_0}{1-y} \; , \label{sigma0strangeness}
\end{align}
where $\sigma_0$ is given by
\begin{align}
\sigma_0 = \frac{\hat{m}}{2 m_N} \bra{N} \bar{u} u + \bar{d} d - 2 \bar{s} s \ket{N} \; . \label{sigma0def}
\end{align}
This $\sigma_0$ is the central  quantity  of the calculations in this paper,
because it allows one to deduce the strangeneness content of the nucleon. For example,
if $y$ is equal to zero, $\sigma_{\pi N}$ and $\sigma_0$ are identical and the nucleon 
has a pure $u$- and $d$-quark content. 

\subsection{Method of calculation}

A possible way to calculate $\sigma_{\pi N}$ is based on utilizing the Feynman-Hellmann theorem.
The starting point is the trace of the energy-momentum tensor of QCD, which  in the isospin limit reads
(neglecting heavy quarks and higher order QCD corrections, for more details we refer e.g.
to Ref.~\cite{Donoghue:1992dd}) 
\begin{equation}
\left( T_{\mathrm{QCD}} \right)_{\mu}{}^{\mu} = \frac{\beta_{\mathrm{QCD}}}{2 g_s} F^{a}_{\mu \nu} F^{a, \, \mu \nu} + \hat{m} \left( \bar{u} u +  \bar{d} d \right) + m_s \bar{s} s \; , 
\label{QCDemt}
\end{equation}
with $\beta_{\mathrm{QCD}}$ the beta-function of QCD. The  expectation value with a nucleon state follows as
\begin{equation}
  \bra{N} \left( T_{\mathrm{QCD}} \right)_{\mu}{}^{\mu} \ket{N} = \bra{N} m_N^2 \ket{N}
  = m_N^2 \braket{N | N} = m_N^2 \; . \label{eq38}
\end{equation}
Further, the nucleon mass $m_N$ is a function of the quark masses $m_N ( \hat{m}, m_s)$. 
Hence we can investigate Eq.~(\ref{eq38}) by taking the derivative with respect to $\hat{m}$,
\begin{equation}
\frac{\partial}{\partial \hat{m}} \left( m_N^2 \right) = 2 m_N \left( \frac{\partial m_N}{\partial \hat{m}} \right) 
= \bra{N} \left( \frac{\partial}{\partial \hat{m}}  \left( T_{\mathrm{QCD}} \right)_{\mu}{}^{\mu} \right) \ket{N} + m_N^2 \frac{\partial}{\partial \hat{m}} \braket{N | N} \; .
\end{equation}
The second term vanishes due to normalization and we are left with
\begin{align}
2 m_N \left( \frac{\partial m_N}{\partial \hat{m}} \right) = \bra{N} \left( \frac{\partial}{\partial \hat{m}}  \left( T_{\mathrm{QCD}} \right)_{\mu}{}^{\mu} \right) \ket{N} \overset{(\ref{QCDemt})}{=} \bra{N} \bar{u} u + \bar{d} d \ket{N} \; .
\end{align} 
From this relation we see that
\begin{align}
\hat{m} \left( \frac{\partial m_N}{\partial \hat{m}} \right) = \frac{\hat{m}}{2 m_N} \bra{N} \bar{u} u + \bar{d} d \ket{N} = \sigma_{\pi N} \; ,
\end{align}
which is known as the Feynman-Hellmann theorem for the pion-nucleon sigma-term. 
A similar calculation, where the nucleon mass is differentiated with respect to the strange quark mass $m_s$,
leads to $\sigma_s$,
\begin{equation}
\sigma_s = m_s \left( \frac{\partial m_N}{\partial m_s} \right) \; . \label{sigmas}
\end{equation}
Thus, the pion-nucleon sigma-term can be calculated if we know the nucleon mass as a function of the quark masses. 
In the continuum, the nucleon mass is  calculable within  chiral perturbation theory (CHPT). 
We will closely follow the procedure shown in Ref.~\cite{Borasoy:1996bx} to obtain the sigma-term and $\sigma_0$. 
Alternatively, one can utilize lattice QCD, but we will not follow that path here as explained
in the introduction.

\subsection{Some phenomenology}

For a long time, the value of $\sigma_{\pi N}$ was taken as $(45\pm5)\,$MeV~\cite{Gasser:1990ce},
but this value is now superseded by the recent Roy-Steiner analysis of pion-nucleon scattering
that also includes the superb measurements from pionic hydrogen and deuterium, leading to
$\sigma_{\pi N} = (59.1 \pm3.5)\,$MeV~\cite{Hoferichter:2015dsa}, for more details, see the 
review~\cite{Hoferichter:2015hva}.
Even if one ignores the constraints from the pionic atom measurements, fitting a representation
based on Roy-Steiner equations to the low-energy pion-nucleon scattering data base leads to a
consistent but less precise value of $(58\pm5)\,$MeV~\cite{RuizdeElvira:2017stg}.

So what do we know about $\sigma_0$? For a long time, the pioneering calculations of
Refs.~\cite{Gasser:1980sb} and \cite{Borasoy:1996bx} led to values of $(35\pm 5)\,$MeV and $(36\pm 7)$~MeV,
respectively, taken together with the old value of  $\sigma_{\pi N}$ suggesting a small
strangeness content. This, however, clearly is at odds with the new value of $\sigma_{\pi N}$.
However, more recent calculations in the heavy baryon as well as covariant scheme with and
without delta contributions to third order gave a vary unclear picture, with central
values of $\sigma_0$ ranging from 46 to 89~MeV, see Tab.~3 of Ref.~\cite{Alarcon:2012nr}, with
disturbingly large differences between the HB and covariant approaches when the decuplet
was included. Clearly, such a situation requires an improved  fourth order calculation,
as will be presented in the following.

\section{Chiral Lagrangians}
\label{sec:formalism}

In what follows, we will utilize baryon chiral perturbation theory (BCHPT) in
various formulations. Here, we briefly exhibit the pertinent chiral Lagrangians.
Since our aim is the calculation of the octet baryon masses to obtain the sigma-terms up to chiral
order $\mathcal{O}(p^4)$, we only consider the effective baryon Lagrangians, which
are relevant for our calculations. As we seek the flavor decomposition of $\sigma_{\pi N}$,
we must work in three-flavor baryon chiral perturbation theory. Note that the
purely mesonic Lagrangian is given in App.~\ref{app:A} together with some definitions of various
basic quantities.

\subsection{Baryon Lagrangians}
The leading-order baryon Lagrangian with coupling to the octet-meson fields is given
by, see e.g.~\cite{Krause:1990xc},
\begin{equation}
  \mathcal{L}_{\phi B}^{(1)} = \text{Tr} \left( \bar{B} \left( i \slashed{D}
  - m_0 \right) B \right) + \frac{D}{2} \text{Tr} \left( \bar{B} \gamma^{\mu} \gamma_5
  \left\lbrace u_{\mu} , B \right\rbrace \right) + \frac{F}{2} \text{Tr} \left(
  \bar{B} \gamma^{\mu} \gamma_5 \left[ u_{\mu} , B \right] \right) \; ,
  \label{MBL1}
\end{equation}
where $m_0$ is the octet-baryon mass in the chiral limit and $B(x)$ is a traceless $3 \times 3$-matrix
denoting the lowest-lying octet-baryon fields in flavor SU(3)
\begin{equation}
	B (x) = \frac{1}{\sqrt{2}} \sum_{a=1}^{8} \lambda^a B^a(x) =  \left(
	\begin{array}{ccc}
	\frac{1}{\sqrt{2}} \Sigma^0 + \frac{1}{\sqrt{6}} \Lambda & \Sigma^+ & p \\
	\Sigma^- & - \frac{1}{\sqrt{2}} \Sigma^0 + \frac{1}{\sqrt{6}}  \Lambda & n \\
	\Xi^- & \Xi^0 &  - \frac{2}{\sqrt{6}}  \Lambda \\
	\end{array}
	\right) \; .
        \label{BaryonMatrix}
\end{equation}
The matrix $B$ transforms as $B \to K B K^{\dagger}$ under $\text{SU}(3)_L \times \text{SU}(3)_R$
transformations, where $K(L,R,U)$ is the so called compensator field. $K$ is an element of the
conserved subgroup $\text{SU}(3)_V$. It depends on the left- and right-handed fields $L$, $R$,
and on the pseudo-Goldstone boson fields collected in $U(x)$, cf. App.~\ref{app:A},
which makes it a local transformation.
The covariant derivative is defined to obey the transformation property $D_{\mu} B \to K
( D_{\mu} B ) K^{\dagger}$ and is given by
\begin{equation}
D_{\mu} B = \partial_{\mu} B + \left[ \Gamma_{\mu}, \, B \right] \; ,
\end{equation} 
with
\begin{equation}
  \Gamma_{\mu} = \frac{1}{2} \left \lbrace u^{\dagger} \left( \partial_{\mu} - i r_{\mu} \right) u
  + u \left( \partial_{\mu} - i l_{\mu} \right) u^{\dagger} \right \rbrace \; ,
\end{equation}
where $u = \sqrt{U} = \exp \left( i \phi / (2 F_{\phi}) \right)$. The chiral vielbein is given by
\begin{align}
u_{\mu} = i \left \lbrace u^{\dagger} \left( \partial_{\mu} - i r_{\mu} \right) u
- u \left( \partial_{\mu} - i l_{\mu} \right) u^{\dagger} \right \rbrace \; , \label{Vielbein}
\end{align}
which also transforms as $u_{\mu} \to K u_{\mu} K^{\dagger}$. As we are only interested in masses
and $\sigma$-terms, we set the external fields $r_{\mu}$ and $l_{\mu}$ to zero. The expansion
of the vielbein is $u_{\mu} = -\partial_{\mu} \phi/F_\phi + \mathcal{O}(\phi^3)$. Here, $\phi$ denotes
the pseudoscalar fields (pseudo-Goldstone bosons) and $F_\phi$ is the decay constant (in the
chiral limit).
The second and third terms in Eq.~(\ref{MBL1}) introduce axial-vector interactions with the axial-vector
coupling constants $D$ and $F$,
which can be determined from semi-leptonic decays. Throughout, we use $D=0.80$ and $F=0.46$, so that
$g_A = F+D= 1.26$, with $g_A$ the nucleon axial-vector coupling. This is the first matching relation between
the two- and three-flavor versions of BCHPT of relevance here.

From a power counting perspective the chiral vielbein contains derivatives of meson octet fields
and counts as $\mathcal{O}(p)$. The octet-baryon mass term $m_0$ has chiral order zero, since it
is of the same order of magnitude as the chiral symmetry breaking scale $\Lambda_{\chi}$
and thus cannot be used as a small expansion parameter. The same argument holds for the baryon
momenta, which are generated by the derivative term $ i \slashed{D} $. The difference
$(i \slashed{D} - m_0)$, however, corresponds to $(\slashed{p}-m_0)$ in momentum space,
which is considered to be small and therefore counts as $\mathcal{O}(p)$. These properties can be used
to set up higher-order baryon Lagrangians. 

The second order baryon Lagrangian includes terms with quark mass insertions that explicitly
break the chiral symmetry, terms with two vielbeins $u_{\mu}$, and terms with external currents
\cite{Frink:2006hx,Oller:2006yh}, see also \cite{Ren:2012aj},
\begin{align}
\mathcal{L}_{\phi B}^{(2)} = \mathcal{L}_{\phi B}^{(2, \, sb.)} + \mathcal{L}_{\phi B}^{(2, \, int.)} \; ,
\end{align}
where the explicit chiral symmetry breaking terms are given by
\begin{align}
\mathcal{L}_{\phi B}^{(2, \, sb.)} = b_0 \text{Tr} \left( \chi_{+}^{} \right) \text{Tr}
\left( \bar{B} B \right) + b_D \text{Tr} \left( \bar{B} \left\lbrace \chi_{+}^{} , B
\right\rbrace \right) + b_F \text{Tr} \left( \bar{B} \left[ \chi_{+}^{} , B \right] \right) \; ,
\label{MBL2sb}
\end{align}
and the $\mathcal{O}(p^2)$ interaction terms by
\begin{align}
\begin{split}
\mathcal{L}_{\phi B}^{(2, \, int.)} = \; & b_1 \text{Tr} \left( \bar{B} \left[ u_{\mu} ,
\left[ u^{\mu}, B \right]  \right] \right) + b_2 \text{Tr} \left( \bar{B} \left\lbrace u_{\mu} ,
\left\lbrace u^{\mu},B \right\rbrace  \right\rbrace \right) \\
& + b_3 \text{Tr} \left( \bar{B} \left\lbrace u_{\mu} , \left[ u^{\mu},B \right]  \right\rbrace \right)
+ b_4 \text{Tr} \left( \bar{B} B \right) \text{Tr} \left( u^{\mu} u_{\mu} \right) \\ 
& + i b_5 \left( \text{Tr} \left( \bar{B} \left[ u^{\mu} , \left[ u^{\nu}, \gamma_{\mu} D_{\nu} B
\right]  \right] \right) - \text{Tr} \left( \bar{B} \overleftarrow{D}_{\nu} \left[ u^{\nu} ,
\left[ u^{\mu}, \gamma_{\mu} B \right]  \right] \right)  \right) \\
& + i b_6 \left( \text{Tr} \left( \bar{B} \left[ u^{\mu} , \left\lbrace u^{\nu}, \gamma_{\mu}
  D_{\nu} B \right\rbrace  \right] \right) - \text{Tr} \left( \bar{B} \overleftarrow{D}_{\nu}
\left\lbrace u^{\nu} , \left[ u^{\mu}, \gamma_{\mu} B \right]  \right\rbrace \right)  \right) \\
& + i b_7 \left( \text{Tr} \left( \bar{B} \left\lbrace u^{\mu} , \left\lbrace u^{\nu},
\gamma_{\mu} D_{\nu} B \right\rbrace  \right\rbrace \right) - \text{Tr} \left( \bar{B}
\overleftarrow{D}_{\nu} \left\lbrace u^{\nu} , \left\lbrace u^{\mu}, \gamma_{\mu} B \right\rbrace
\right\rbrace \right)  \right) \\
& + i b_8 \left( \text{Tr} \left( \bar{B} \gamma_{\mu} D_{\nu} B \right) - \text{Tr}
\left( \bar{B} \overleftarrow{D}_{\nu} \gamma_{\mu} B \right) \right) \text{Tr}
\left( u^{\mu} u^{\nu}  \right) + \ldots \; ,
\end{split} \label{MBL2int}
\end{align}
where $b_0$, $b_D$, $b_F$, and $b_{1,2,...,8}$ are LECs and $\chi_{+}^{} = u^{\dagger} \chi u^{\dagger}
+ u \chi^{\dagger} u$. Note that $b_0$, $b_D$, $b_F$, and $b_{1,...,4}$ have dimension
$(\text{mass})^{-1}$ and $b_{5,...,8}$ have $(\text{mass})^{-2}$. The ellipses in Eq.~(\ref{MBL2int})
denotes terms that do not contribute to the calculation of the baryon masses up to order $\mathcal{O}(p^4)$.
The third order meson-baryon Lagrangian does not contribute to the masses, because it generates
meson-baryon interactions that will not enter before $\mathcal{O}(p^5)$. The fourth order Lagrangian,
however, will contribute via tree level diagrams. The relevant part is given by 
\begin{align}
\begin{split}
\mathcal{L}_{\phi B}^{(4)} = \; & d_1 \text{Tr} \left( \bar{B} \left[ \chi_{+}^{} ,
\left[\chi_{+}^{} , B \right]  \right] \right) + d_2 \text{Tr} \left( \bar{B} \left[ \chi_{+}^{} ,
\left\lbrace \chi_{+}^{} , B \right\rbrace  \right] \right) \\
& + d_3 \text{Tr} \left( \bar{B} \left\lbrace \chi_{+}^{} , \left\lbrace \chi_{+}^{} ,B \right\rbrace
\right\rbrace \right) + d_4 \text{Tr} \left( \bar{B} \chi_{+}^{} \right) \text{Tr} \left(  \chi_{+}^{} B
\right) \\
& + d_5 \text{Tr} \left( \bar{B} \left[ \chi_{+}^{} , B \right] \right) \text{Tr} \left(  \chi_{+}^{}
\right) + d_7 \text{Tr} \left( \bar{B} B \right) \left[ \text{Tr} \left( \chi_{+}^{} \right) \right]^2 \\
& + d_8 \text{Tr} \left( \bar{B} B \right) \text{Tr} \left( \chi_{+}^2 \right) \; ,
\label{MBL4}
\end{split}
\end{align} 
where $d_{1,...,5}$, $d_{7}$, and $d_8$ are LECs as well with dimension $(\text{mass})^{-3}$. 

\subsection{The heavy-baryon approach} \label{SecHBapproach}

We have seen that the baryon Lagrangian at lowest order introduces a new parameter $m_0$,
which is close to the chiral symmetry breaking scale and does not vanish in the chiral limit. 
It was first pointed out in~\cite{Gasser:1987rb} that this spoils
the power counting scheme in loop calculations when using the relativistic nucleon
propagator. In order to restore the power counting, the so-called heavy-baryon approach (HB)
was introduced in Ref.~\cite{Jenkins:1990jv} and systematically developed in Ref.~\cite{Bernard:1992qa}.
The baryons are considered as  very heavy sources with momentum
\begin{align}
p_{\mu} = m_0 v_{\mu} + l_{\mu} \; , \label{nonrelmom}
\end{align}
where $v_{\mu}$ is the four-velocity subject to the constraint $v^2 = 1$, and $l_{\mu}$
is a small off-shell momentum, $v \cdot l \ll m_0 $. The baryon field $B$ can then be written as
\begin{align}
B = \exp\{-i m_0 v \cdot x\}  \left(  B_v + b_v \right) \; ,
  \label{Bvdecomp}
\end{align}
with a large component field $B_v$ and a small component field $b_v$, satisfying $\slashed{v} B_v = B_v$
and $ \slashed{v} b_v = - b_v$. Inserting this into Eq.~(\ref{MBL1}) one  obtains a Lagrangian in terms
of $B_v$. Within the path integral formalism, one can shift the variable $b_v$ to absorb mixing
terms of $b_v$ and $B_v$. Afterwards the $b_v$ field is integrated out. The result is given by
\begin{align}
\mathcal{L}_{HB}^{(1)} = \text{Tr} \left( \bar{B}_v \left( i v \cdot D \right) B_v \right) + D \, \text{Tr} \left( \bar{B}_v S^{\mu} \left\lbrace u_{\mu} , B_v \right\rbrace \right) + F \, \text{Tr} \left( \bar{B}_v S^{\mu} \left[ u_{\mu} , B_v \right] \right) \; , \label{HBL1}
\end{align}
plus terms of $\mathcal{O}(1/m_0)$. $S^{\mu}$ is the covariant Pauli-Lubanski spin operator defined by
$S^{\mu} = -\gamma_5 \left( \gamma^{\mu} \slashed{v} - v^{\mu} \right)/2$, 
with $v \cdot S = 0$ and $S^2 = (1-D)/4$ in $D$ space-time dimensions. 
The HB Lagrangian does not contain a mass term for $B_v$ and its corresponding propagator is 
\begin{align}
  S^{ab}_{HB}(\omega) = \frac{i \delta^{a b}}{v \cdot k + i \epsilon} \; ,
  \quad \text{with} \; \; \omega = v \cdot k \; . \label{NHBprop}
\end{align}
Using this heavy-baryon propagator in loop calculations restores the power counting,
since the mass parameter $m_0$ does not appear. The problem with the HB approach, however, is
that $m_0$ is  not extremely large and in some calculations one expects significant corrections
from $\mathcal{O}(1/m_0)$ terms. For more details on this and also the representation of the
effective Lagrangian in this basis, see the reviews~\cite{Bernard:1995dp,Bernard:2007zu}.

\subsection{The EOMS scheme} \label{SecEOMSscheme}

Despite the fact that the heavy-baryon approach provides a good approximation for many
calculations and restores the power counting, one was still interested to use the fully
covariant meson-baryon Lagrangian. The idea was to use a different renormalization procedure,
rather than the $\overline{\text{MS}}$ and $\widetilde{\text{MS}}$ schemes, which are commonly
used in loop calculations. $\overline{\text{MS}}$ and $\widetilde{\text{MS}}$ use redefinitions of
the parameters in the Lagrangian to subtract the infinities, that arise from the loop diagrams.
The  $\widetilde{\text{MS}}$ scheme is commonly used in CHPT calculations.
The so-called extended-on-mass-shell (EOMS) renormalization scheme \cite{Fuchs:2003qc}, which is
nowadays often used in BCHPT  calculations, achieves this. Within the EOMS scheme, one performs
additional finite subtractions
to cancel the power counting violating terms, i.e. the LECs in the Lagrangian absorb the infinities
and the finite power counting breaking pieces. This ensures that a given diagram will not contribute
to orders lower than its chiral dimension and therefore restores the power counting. The power counting
violating terms can be found by expanding the loop functions in terms of quantities with a known
chiral order, like $M_{\phi}$ ($\phi = \pi, K, \eta$), $(\slashed{p}-m_0)$ or $(p^2 - m_0^2)$. All
terms with a lower chiral dimension than the diagram are then subtracted. Note that the proper
matching of the EOMS scheme to the HB approach is discussed in~\cite{Siemens:2016hdi}. We follow
that paper in our work.

We will use the EOMS scheme (with $\widetilde{\text{MS}}$) and the HB approach for the calculation
of the baryon masses and compare the results. Explicit formulas will be given in the respective sections.

\subsection{Inclusion of the decuplet baryon resonances}
\label{SecInclusionDecuplet}
So far, we only considered Lagrangians describing the octet-baryon fields, the octet-meson fields
and their interactions. However, it was argued early in~\cite{Jenkins:1991es} that
the lowest-lying spin-$3/2$ decuplet-baryon resonances can contribute significantly to the
quantum corrections of the
octet baryon observables. The average octet mass $\bar{m}_B$ and the average decuplet mass $\bar{m}_D$
are only separated by approximately $\bar{m}_D - \bar{m}_B \simeq {231}\,$MeV \cite{Bernard:1993nj}.
This separation is smaller than the $K$ or $\eta$ masses. Further, the coupling $g_{\pi N \Delta}$
between the delta-baryons, nucleons and pions is quite large, see e.g~\cite{Bernard:1996gq}, so
one expects notable effects from the decuplet. 

\subsubsection{Covariant Lagrangian}
The spin-$3/2$ decuplet-baryons are described by the Rarita-Schwinger fields $T_{\mu}^{abc}$.
We use the conventions from~\cite{Bernard:1995dp}, where the $u$, $d$ and $s$ quarks are assigned
the values $1$, $2$ and $3$, respectively, and $a,b,c \in \lbrace 1,2,3 \rbrace$. The decuplet
fields are defined by
\begin{eqnarray}
T_{\mu}^{111} &=& \Delta_{\mu}^{++} \; , \; T_{\mu}^{112}= T_{\mu}^{121} = T_{\mu}^{211} =
\frac{\Delta_{\mu}^{+}}{\sqrt{3}} \; , \; T_{\mu}^{122}= T_{\mu}^{212} = T_{\mu}^{221} =
\frac{\Delta_{\mu}^{0}}{\sqrt{3}}  \; , \; T_{\mu}^{222} = \Delta_{\mu}^{-} \; , \nn
T_{\mu}^{113}&=& T_{\mu}^{131} = T_{\mu}^{311} = \frac{\Sigma_{\mu}^{*+}}{\sqrt{3}} \; , \; T_{\mu}^{223}
= T_{\mu}^{232} = T_{\mu}^{322} = \frac{\Sigma_{\mu}^{*-}}{\sqrt{3}} \; , \nn
T_{\mu}^{123} &=& T_{\mu}^{132} = T_{\mu}^{213} = T_{\mu}^{231} = T_{\mu}^{312} = T_{\mu}^{321} =
\frac{\Sigma_{\mu}^{*0}}{\sqrt{6}} \; , \\
T_{\mu}^{133}&=& T_{\mu}^{313} = T_{\mu}^{331} = \frac{\Xi_{\mu}^{*0}}{\sqrt{3}} \; , \; T_{\mu}^{233}
= T_{\mu}^{323} = T_{\mu}^{332} = \frac{\Xi_{\mu}^{*-}}{\sqrt{3}} \; , \; T_{\mu}^{333} = \Omega_{\mu}^{-} \; .
\nonumber
\label{Decupletfields}
\end{eqnarray}
$T_{\mu}^{abc}$ is totally symmetric under permutations of $a$, $b$ and $c$. The first order
decuplet Lagrangian reads 
\begin{align}
\mathcal{L}_{D}^{(1)} = \bar{T}_{\mu}^{abc} \left( i \gamma^{\mu \nu \rho} D_{\rho} - m_D
\gamma^{\mu \nu} \right) T_{\nu, \, abc} \; , \label{DecL1}
\end{align}
where $m_D$ is the decuplet-baryon mass in the chiral limit, $\gamma^{\mu \nu} := \gamma^{\mu}
\gamma^{\nu} - g^{\mu \nu}$, and $\gamma^{\mu \nu \rho} = (1/2) \lbrace \gamma^{\mu \nu} , \gamma^{\rho}
\rbrace$. The covariant derivative is of the same form as in the meson-baryon Lagrangian and given by 
\begin{align}
D_{\rho} T_{\nu, \, abc} := \partial_{\rho} T_{\nu, \, abc} + \left( \Gamma_{\rho}, T_{\nu} \right)_{abc}  \; , 
\end{align}
with 
\begin{align}
\left( \Gamma_{\rho}, T_{\nu} \right)_{abc} := \left( \Gamma_{\rho} \right)^{d}_{a}
T_{\nu, \, dbc} + \left( \Gamma_{\rho} \right)^{d}_{b} T_{\nu, \, adc} +
\left( \Gamma_{\rho} \right)^{d}_{c} T_{\nu, \, abd} \; , 
\end{align}
where $\left( \Gamma_{\rho} \right)^{d}_{a}$ denotes the element in row $a$ and column $d$ of
the chiral connection $\Gamma_{\rho}$. The relativistic spin-$3/2$ propagator in $D$ space-time
dimensions takes the form 
\begin{align}
G^{\rho \mu} (k) =  \frac{-i (\slashed{k} + m_D^{} )}{k^2 - m_D^2 + i \epsilon} \left( g^{ \rho \mu }
- \frac{1}{D-1} \gamma^{\rho} \gamma^{\mu} + \frac{k^{\rho}  \gamma^{\mu} -  \gamma^{\rho} k^{\mu}}
{(D-1) m_D^{} } - \frac{D-2}{(D-1) m_D^{2} } k^{\rho} k^{\mu} \right) \; .
\label{decuprop}
\end{align}
Due to the non-vanishing quark masses, there is also a second order symmetry breaking Lagrangian
\begin{align}
\mathcal{L}_{D}^{(2, \, sb.)} = \frac{t_0}{2} \text{Tr} \left( \chi_{+}^{} \right)  \bar{T}_{\mu}^{abc}
g^{\mu \nu} T_{\nu, \, abc} + \frac{t_D}{2} \bar{T}_{\mu}^{abc} g^{\mu \nu} \left( \chi_{+}^{} ,
T_{\nu} \right)_{abc} \; ,
\label{DecL2sb}
\end{align}
with the LECs $t_0$ and $t_D$. The leading-order interaction between the decuplet fields, the
octet baryons and the mesons of chiral order one is usually written as 
\begin{align}
\mathcal{L}_{D B \phi}^{(1)} = \frac{\mathcal{C}}{2} \left\lbrace \bar{T}_{\mu}^{abc} \,
\Theta^{\mu \nu}(z) \left( u_{\nu} \right)_{a}^{i} B_{b}^{j} \epsilon_{c i j} + \bar{B}_{j}^{b}
\left( u_{\nu} \right)_{i}^{a} \Theta^{\nu \mu}(z) T_{\mu, \, abc} \epsilon^{c i j} \right\rbrace \; ,
\label{DBMint}
\end{align}
where $C$ is a coupling constant, which can for example be fixed from the decay $\Delta \to N \pi $,
$ \epsilon_{c i j}$ is the Levi-Civita tensor and 
\begin{align}
\Theta^{\mu \nu}(z) = g^{\mu \nu} - \left( z + \frac{1}{2} \right) \gamma^{\mu} \gamma^{\nu} \; ,
\end{align}
with the so-called off-shell parameter $z$ describing the coupling of the ``off-shell'' spin-$1/2$
components from the Rarita-Schwinger field. As shown in Ref.~\cite{Krebs:2009bf}, $z$ can be absorbed into
redefinitions of certain LECs and is therefore redundant, see also Ref.~\cite{Tang:1996sq}.
Nevertheless, we will use the value $z=1/2$ in our covariant calculations as done in Ref.~\cite{Yao:2016vbz},
see also Ref.~\cite{Wies:2006rv}.

\subsubsection{HB approach for the decuplet}
Since the decuplet-baryons also possess a large mass, it is quite obvious to apply the heavy-baryon approximation to them as well, for a detailed discussion see~\cite{Hemmert:1997ye}.
Analogous to the nucleon case, one  expresses the decuplet fields in terms of velocity eigenstates,
\begin{align}
  T_{\mu}^{abc} = e^{-i m_0 v \cdot x} \left\lbrace \left( T_v \right)_{\mu}^{abc}
  + \left( t_v \right)_{\mu}^{abc} \right\rbrace \; .
\end{align} 
Note that the exponential function contains the average octet-baryon mass in the chiral limit,
$m_0$, and not the decuplet mass $m_D$. This is done to avoid complex exponential functions in
the decuplet-octet-meson interaction term. The price we have to pay is that the mass parameter
$m_D$ will not cancel completely in the Lagrangian. Instead, the HB Lagrangian maintains a mass
scale $\Delta := m_D - m_0 $, which does not vanish in the chiral limit and counts as
order $\mathcal{O}(p)$ within the power-counting scheme. It takes the form 
\begin{align}
\begin{split}
  \mathcal{L}_{DHB}^{(1)} = & -i \bar{T}_{\mu}^{abc} ( v \cdot D ) T_{abc}^{\mu} - \Delta
  \bar{T}_{\mu}^{abc} T_{abc}^{\mu} \\ 
  & + \frac{\mathcal{C}}{2} \left\lbrace \bar{T}_{\mu}^{abc} \left( u^{\mu} \right)_{a}^{i} B_{b}^{j}
  \epsilon_{c i j} + \bar{B}_{j}^{b} \left( u_{\mu} \right)_{i}^{a} T_{abc}^{\mu} \epsilon^{c i j}
  \right\rbrace \; .
  \label{DeltaL1HB}
\end{split}
\end{align}
In the HB approach the off-shell parameter is usually set to $-1/2$, so that $ \Theta^{\mu \nu} $
reduces to the Minkowski metric $g^{\mu \nu}$. The propagator of the decuplet field (in $D$ dimensions)
simplifies to 
\begin{align}
  G_{HB}^{\mu \nu} (k) = \frac{i P^{\mu \nu} }{v \cdot k - \Delta + i \epsilon} \; , \quad \text{with} \;
  \; P^{\mu \nu} = v^{\mu} v^{\nu} - g^{\mu \nu} - 4 \left( \frac{D-3}{D-1} \right) S^{\mu} S^{\nu} \; ,
  \label{DecPropHB}
\end{align}
containing explicitly the scale $\Delta$. The projection operator $P^{\mu \nu}$ satisfies
$v_{\mu} P^{\mu \nu} = P^{\mu \nu} v_{\nu} = 0$ and $P^{\mu}{}_{\mu} = -2$. 

\subsection{Constraints on the LECs}
\label{sec:constraints}

From the above it is clear that the baryon masses are not sufficient to fix all the
LECs that appear. In fact, this does not appear possible within the continuum. So
one way would be to resort to lattice QCD, which allows to vary the quark masses
and thus gives a better handle on the symmetry breaking LECs, see
e.g.
Refs.~\cite{Ren:2012aj,Young:2002ib,Bernard:2003rp,Procura:2003ig,Beane:2004ks,Frink:2005ru,WalkerLoud:2011ab,Semke:2012gs,Lutz:2018cqo}.
However, there is an unresolved
discrepancy between the precise Roy-Steiner determination of $\sigma_{\pi N}$
and present lattice QCD calculations, see Ref.~\cite{Hoferichter:2016ocj} (and references therein).
Therefore, we follow another path and try to constrain the LECs (or combinations thereof)
as much as possible utilizing continuum data. We consider matching between the
SU(3) and SU(2) versions of the effective field theory, which gives relations
between the coupling constants. More precisely, integrating out the strange quark
reduces  three-flavor CHPT to the two-flavor theory. This program has been carried
out in detail in Refs.~\cite{Frink:2004ic,Mai:2009ce}. We will use the matching
relations given in Eqs.~(5.4,5.5,5.6) of Ref.~\cite{Frink:2004ic} because these
also include some information on the fourth order LECs $d_i$.
The dimension-two SU(2) LECs $c_{1,2,3,4}$ have been most precisely  determined
from matching the Roy-Steiner analysis of pion-nucleon scattering to the CHPT amplitudes
\cite{Hoferichter:2015tha}. We use the values obtained in the standard power counting
from that paper,
\begin{equation}
  c_1 = -1.11(3)~, ~~
  c_2 =  3.13(3)~, ~~
  c_3 = -5.61(6)~, ~~
  c_4 =  4.26(4)~,
\end{equation}  
all in GeV$^{-1}$. Note that we will not use the matching relation of $c_4$ as
it involves dimension-two LECs that do  not appear in the baryon masses, see Ref.~\cite{Mai:2009ce}.
When the decuplet is included, the values of the $c_i$ are changed as the $\Delta (1232)$-contribution
has to be subtracted. We follow Ref.~\cite{Bernard:1996gq}, adopting to the value of $g_A$ used here.
The $\Delta (1232)$-contribution is given by 
\begin{align}
c_2^{\Delta} = -c_3^{\Delta} 
= \frac{g_A^2 \left( m_{\Delta} - m_N \right) }{2 \left[ ( m_{\Delta} - m_N )^2 - M_{\pi}^2 \right]} = 3.49~{\rm GeV}^{-1}\; , 
\end{align}
with $m_{\Delta}$ and $m_{N}$ the average delta and nucleon mass, respectively. 
Note that the value of $c_1$ is assumed to be generated from scalar sources only~\cite{Bernard:1996gq}.
The matching of the $c_i$ and other LECs between the deltafull and deltaless theory has
been refined in~\cite{Siemens:2016jwj}.
Alternatively to constraining the $c_i$, one could match to the scattering length expressions
given in \cite{Mai:2009ce}, see also \cite{Oller:2006jw}, but since the corresponding
calculations are not available in the EOMS scheme to sufficient accuracy, we do not follow this path here.

\section{Calculation of the baryon masses and the $\sigma$-term}
\label{SecMassCalc}
Now we have all the information that we need to calculate the quantum corrections of the
octet-baryon masses. These corrections are given by the baryon self-energy $\Sigma_{B} $,
which can be determined from the one-particle-irreducible perturbative contribution to
the two-point function of the baryon field $B$
\begin{align}
i S_B (\slashed{p}) := \frac{i}{\slashed{p} - m_0 - \Sigma_{B}(\slashed{p}) } \; .
\end{align}
The physical baryon mass $m_B$ is the pole of the propagator at
$\slashed{p}=m_B$, i.e. 
\begin{align}
m_B - m_0 - \Sigma_{B}( \slashed{p}=m_B ) = 0 \quad \Rightarrow
\quad m_B =  m_0 + \Sigma_{B}( \slashed{p}=m_B ) \; .
\end{align}
Since we do not have the exact form of $m_B$ and we only calculate the mass up to a specific
order, we can only approximate the self-energy by setting $\slashed{p}= m_0 +
(\text{h.o.c.})$\footnote{Here, h.o.c. means higher-order corrections. We have to adjust the value
  of $\slashed{p}$ depending on the accuracy of our calculation.}. 
In the heavy baryon approach the propagator has a slightly different form, cf. Eq.~(\ref{NHBprop}).
The baryon mass is given by 
\begin{align}
m_B = m_0 + \Sigma_{HB}(\omega = 0) \; , \label{HBselfenergyCalc}
\end{align}
where $\Sigma_{HB}$ is the self-energy in the HB approach and $\omega=v \cdot p$. 

\begin{figure}[t]
\begin{center}  
\includegraphics[width=0.60\linewidth]{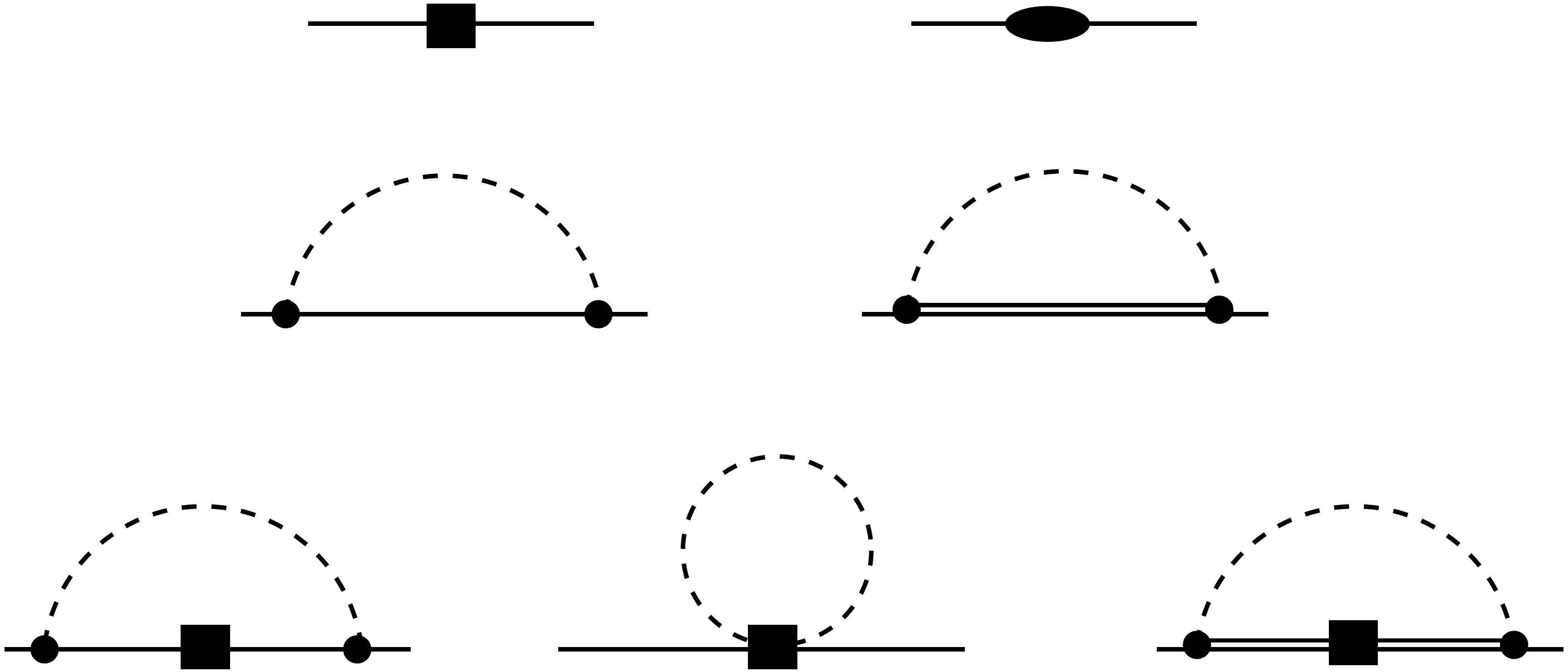}
\end{center}
\caption{
Feynman diagrams contributing to the masses  up to fourth order. The contact terms at 
second and fourth order, the leading loop and the  next-to-leading
loop corrections are given in the first, second and third row, respectively.
Solid, dashed, and solid double lines refer to octet baryons,
Goldstone bosons, and decuplet states, respectively.
Vertices denoted by a filled dot/square/ellipse refer to insertions from the
first/second/fourth order chiral Lagrangian, respectively.
Diagrams contributing via wave function renormalization only are not shown.
\label{fig:diags}}
\end{figure}

It is important to note that the self-energy -- in the covariant as well as in the HB formulation --
is a matrix $\Sigma^{ba}$, depending on the incoming baryon flavor index $a$ and the outgoing
$b$ ($a,b =1,2,...,8$). The self-energies of the octet baryons ($N, \Sigma, \Lambda, \Xi $)
can be calculated by the following linear combinations \cite{Lehnhart:2004vi}
\begin{align}
\Sigma_{N} = \Sigma^{44} - i \Sigma^{54} \; , 
\Sigma_{\Sigma} = \Sigma^{33} \; , 
\Sigma_{\Lambda} = \Sigma^{88} \; , 
\Sigma_{\Xi} = \Sigma^{44} + i \Sigma^{54} \; .
\label{SelfEnergiesB}
\end{align}
In order to calculate the masses/self-energies up to $\mathcal{O}(p^4)$, we need to consider
all relevant terms of the effective Lagrangian
\begin{align}
\mathcal{L}_{\text{eff}}^{} = \mathcal{L}_{\phi B}^{(1)} + \mathcal{L}_{\phi B}^{(2)} + \mathcal{L}_{\phi B}^{(4)} + \mathcal{L}_{\phi}^{(2)} + \mathcal{L}_{\phi}^{(4)} + \mathcal{L}_{D}^{(1)} + \mathcal{L}_{D}^{(2)} \; .
\end{align}
The various contributions to the self-energy at second, third and fourth order are depicted in
Fig.~\ref{fig:diags}. The contributions from these orders will now be discussed separately.

\subsection{Second order}

The calculation of the tree graphs at $\Order(p^2)$ is straightforward and well
documented in the literature, cf. Fig.~\ref{fig:diags} (left diagram in the first row).
The nucleon mass is given by (in both the HB and the EOMS scheme)
\beq
m_N = m_0 - (2 b_0 + 4 b_F) M_{\pi}^2 - (4 b_0 + 4 b_D - 4 b_F) M_{K}^2  ~,
\eeq
and the octet-baryon masses fulfill the Gell-Mann--Okubo relation,
\begin{align}
2 \left( m_N + m_{\Xi} \right) = m_{\Sigma} + 3 m_{\Lambda} \; ,
\end{align}
which turns out to be approximately fulfilled in nature. For second and third order calculations,
we use the mesonic Gell-Mann--Okubo relation
\beq
3M_\eta^2 = 4M_K^2 + M_\pi^2~,
\eeq
which is also  approximately fulfilled in nature. For the sigma-term and
its flavor singlet, we find
\beq
\sigma_{\pi N} = -2 M_{\pi}^2 ( 2 b_0 + b_D + b_F ) \; , \quad \sigma_{0} = 2 M_{\pi}^2 ( b_D - 3 b_F ) \; .
\eeq
As expected,  the sigma-terms vanish in the chiral limit and they depend on the symmetry-breaking
LECs. At this order, these LECs can be fixed from the baryon masses, leading to
\begin{align}
  \sigma_0 = \frac{1}{2} \left( \frac{M_{\pi}^2}{M_{K}^2 - M_{\pi}^2} \right)
  \left( m_{\Xi} + m_{\Sigma} - 2 m_{N} \right) \; .
\label{si0treelevel}
\end{align}
Using the average masses for the corresponding meson and baryon isospin multiplets, one obtains
\begin{align}
\sigma_{0} \simeq {27}\,{\rm MeV} \; .
\end{align}
Taking  $\sigma_{\pi N} = {59.1}\,$MeV face value, this would lead to a strangeness content of
$y\simeq 0.54$, which appears unacceptably large.

\subsection{Third order}

At third order, the baryon mass takes the generic form
\beq
m_B^{} = m_0^{} + m_B^{(2)} + m_{B}^{(3)} + \delta m_{B}^{(3)}~, 
\eeq
with $ m_{B}^{(3)}$ the contribution from the leading one-loop diagrams with octet-baryon
intermediate states and $\delta m_{B}^{(3)}$ the corresponding correction from the decuplet,
cf. Fig.~\ref{fig:diags} (second row). As for the heavy baryon approach, the corresponding
formulas can be found in Ref.~\cite{Bernard:1993nj}, with which we agree. For the theory
without the decuplet, the EOMS expressions have been given first in Ref.~\cite{Lehnhart:2004vi}.
For the decuplet contribution within the EOMS scheme, we find (for a general value of the
off-shell parameter $z$ and the renormalization scale set to $\mu=m_D$):
\begin{align}
\delta m_B^{(3)} =  \frac{C^2}{48 \pi^2 F_{\phi}^2} \left[ \beta_{B \pi}^{(3)} \tilde{H}_{cov}(M_{\pi}) + \beta_{B K}^{(3)} \tilde{H}_{cov}(M_{K}) + \beta_{B \eta}^{(3)} \tilde{H}_{cov}(M_{\eta}) \right] \; ,
\end{align}
with 
\begin{eqnarray}
  \tilde{H}_{cov}(M_{\phi}) &=& \frac{1}{96 m_0^3 m_D^2} \left\lbrace m_0^2 M_{\phi}^2
  \left[6 m_0^4-2 m_0^2 \left(6 m_D^2+M_{\phi}^2 \left(5 z^2+4 z+2\right)\right) \right. \phantom{\sqrt{\frac{P_d^2}{N_d^2}}}  \right. \nn 
&\phantom{=}& \left. \left. +4 m_0^{} m_D^{} M_{\phi}^2 \left(2 z^2-2 z-1\right)+3 \left(2 m_D^4-3 m_D^2 M_{\phi}^2+M_{\phi}^4\right)\right] \phantom{\sqrt{\frac{P_d^2}{N_d^2}}} \right. \nn
&\phantom{=}& \left. -3 m_D^{} M_{\phi}^{} \left(m_0^2+2 m_0^{} m_D^{}+m_D^2-M_{\phi}^2\right)^2 \left(M_{\phi}^2-(m_0^{}-m_D^{})^2\right) \phantom{\sqrt{\frac{P_d^2}{N_d^2}}} \right. \nn
&\phantom{=}& \left. \times \sqrt{\frac{- 4 m_D^2 M_{\phi}^2 + (-m_0^2+m_D^2+M_{\phi}^2)_{\phantom{d}}^2}{m_D^2 M_{\phi}^2}} \right. \nn
&\phantom{=}& \left. \times \log \left( \frac{-m_0^2+m_D^2+M_{\phi}^2}{2 m_D^{} M_{\phi}^{}} - \sqrt{ \left( \frac{-m_0^2+m_D^2+M_{\phi}^2}{2 m_D^{} M_{\phi}^{}} \right)^2 -1 } \right)  \right. \nn 
&\phantom{=}& \left. -3 \left[ -m_0^8-2 m_0^7 m_D^{}+2 m_0^{} m_D^{} \left(m_D^2-M_{\phi}^2\right)^3+\left(m_D^2-M_{\phi}^2\right)^4 \right.  \phantom{\sqrt{\frac{P_d^2}{N_d^2}}} \right. \nn
&\phantom{=}& \left. \left. +6 m_0^5 m_D^{} \left(m_D^2 + M_{\phi}^2\right) + 2 m_0^6 \left(m_D^2 + 2 M_{\phi}^2\right) - 2 m_0^2 \left(m_D^6 - 3 m_D^2 M_{\phi}^4 + 2 M_{\phi}^6\right)^{\phantom{4}} \right. \phantom{\sqrt{\frac{P_d^2}{N_d^2}}} \right. \nn
&\phantom{=}& \left. \left. +4 m_0^4 M_{\phi}^4 \left(2 z^2+4 z-1\right)-2 m_0^3 m_D^{} \left(3 m_D^4+M_{\phi}^4 \left(-16 z^2-8 z+5\right)\right)^{\phantom{4}} \right] \right. \nn
&\phantom{=}& \left. \times \log \left( \frac{M_{\phi}^{}}{m_D^{}} \right)^{\phantom{4}} \right\rbrace \; , \quad \text{for} \; M_{\phi} = M_{\pi} \; .
\end{eqnarray}
and 
\begin{align}
\begin{split}
\tilde{H}_{cov}(M_{\phi}) &= \frac{1}{96 m_0^3 m_D^2} \left\lbrace m_0^2 M_{\phi}^2 \left[6 m_0^4-2 m_0^2 \left(6 m_D^2+M_{\phi}^2 \left(5 z^2+4 z+2\right)\right) \right. \phantom{\sqrt{\frac{P_d^2}{N_d^2}}}  \right. \\ 
& \phantom{=} \left. \left. +4 m_0^{} m_D^{} M_{\phi}^2 \left(2 z^2-2 z-1\right)+3 \left(2 m_D^4-3 m_D^2 M_{\phi}^2+M_{\phi}^4\right)\right] \phantom{\sqrt{\frac{P_d^2}{N_d^2}}} \right. \\
& \phantom{=} \left. -3 m_D^{} M_{\phi}^{} \left(m_0^2+2 m_0^{} m_D^{}+m_D^2-M_{\phi}^2\right)^2 \left(M_{\phi}^2-(m_0^{}-m_D^{})^2\right) \phantom{\sqrt{\frac{P_d^2}{N_d^2}}} \right. \\
  & \phantom{=} \left. \times \sqrt{\frac{4 m_D^2 M_{\phi}^2-(-m_0^2+m_D^2+M_{\phi}^2)_{\phantom{d}}^2}{m_D^2 M_{\phi}^2}} \arccos \left(\frac{-m_0^2+m_D^2+M_{\phi}^2}{2 m_D^{} M_{\phi}^{}}\right)  \right. \nonumber
 \end{split}
\end{align}
\begin{align}
\begin{split} 
& \phantom{=} \left. -3 \left[ -m_0^8-2 m_0^7 m_D^{}+2 m_0^{} m_D^{} \left(m_D^2-M_{\phi}^2\right)^3+\left(m_D^2-M_{\phi}^2\right)^4 \right.  \phantom{\sqrt{\frac{P_d^2}{N_d^2}}} \right. \\
& \phantom{=} \left. \left. +6 m_0^5 m_D^{} \left(m_D^2 + M_{\phi}^2\right) + 2 m_0^6 \left(m_D^2 + 2 M_{\phi}^2\right) - 2 m_0^2 \left(m_D^6 - 3 m_D^2 M_{\phi}^4 + 2 M_{\phi}^6\right)^{\phantom{4}} \right. \phantom{\sqrt{\frac{P_d^2}{N_d^2}}} \right. \\ 
& \phantom{=} \left. \left. +4 m_0^4 M_{\phi}^4 \left(2 z^2+4 z-1\right)-2 m_0^3 m_D^{} \left(3 m_D^4+M_{\phi}^4 \left(-16 z^2-8 z+5\right)\right)^{\phantom{4}} \right] \right. \\
& \phantom{=} \left. \times \log \left( \frac{M_{\phi}^{}}{m_D^{}} \right)^{\phantom{4}} \right\rbrace \; , \quad \text{for} \; M_{\phi} = M_K, M_{\eta} \; ,
\end{split}
\end{align}
and the prefactors $\beta^{(3)}_{B\phi}$ are collected in Tab.~\ref{Op3DecCoeff}, see
also~\cite{Bernard:1993nj}.
The expressions within the curly brackets agree with the ones in Ref.~\cite{Yao:2016vbz} for the
specific choice $z=1/2$ taken there. Note, however, that some of the prefactors are mistyped
in that reference.
\begin{table}[t]
	\centering
	\begin{tabular}{|c|c|c|c|c|} \hline
		& N & $\Sigma$ & $\Lambda$ & $\Xi$ \\ \hline
		$\beta_{B \pi}^{(3)}$ & $8$ & $\frac{4}{3}$ & $6$ & $2$ \\ \hline
		$\beta_{B K}^{(3)}$ & $2$ & $\frac{20}{3}$ & $4$ & $6$  \\ \hline 
		$\beta_{B \eta}^{(3)}$ & $0$ & $2$ & $0$ & $2$ \\ \hline 
	\end{tabular}
	\caption{Coefficients of the $\mathcal{O}(p^3)$ self-energy diagrams with the decuplet-baryon propagator.}
	\label{Op3DecCoeff}
\end{table}

\subsection{Fourth order}

\subsubsection{Baryon masses}
To start this section, we require the fourth order representation of the
Goldstone boson masses. These have been given in the seminal paper \cite{Gasser:1984gg}
and will not be displayed here.

As concerns  the baryon masses, we consider first the tree graphs at $\Order(p^4)$,
see the right diagram in the first row of Fig.~\ref{fig:diags}. Their contribution
is readily evaluated as 
\begin{align}
  m_{B, \, c}^{(4)} = \gamma_{B \pi}^{(4)} M_{\pi}^4 + \gamma_{B K}^{(4)} M_{K}^4
  + \gamma_{B \pi K}^{(4)} M_{\pi}^2 M_{K}^2 \; ,
  \label{ContactIntOp4}
\end{align}
where the coefficients can be found in Tab.~\ref{Table4}. Note that this result is valid for
the HB and the covariant calculation.
\begin{table}[t]
\centering
\resizebox{15cm}{!}{
\begin{tabular}{|c|c|c|c|c|} \hline
	& N & $\Sigma$ & $\Lambda$ & $\Xi$ \\ \hline
$\gamma_{B \pi}^{(4)}$ & $-4\left(4d_1 + 2d_5 + d_7 + 3d_8 \right)$ & $-4\left(4d_3 + d_7 + 3d_8\right) $ & $-4 \left( 4d_3 + \frac{8}{3}d_4 + d_7 + 3d_8 \right) $ & $-4\left(4d_1 - 2d_5 + d_7 + 3d_8 \right) $  \\
		                         & $\phantom{==} $ & $\phantom{=} $ & $\phantom{=}$ & $\phantom{==}$ \\ \hline
$\gamma_{B K}^{(4)}$ & $-16 \left(d_1 - d_2 + d_3 \right.$ & $-16(d_7 + d_8)$ & $-16\left( \frac{8}{3} d_3 + \frac{2}{3} d_4 + d_7 + d_8 \right)$ & $-16 \left(d_1 + d_2 + d_3 \right.$ \\
								&  $\phantom{==} \left. -d_5 + d_7 + d_8 \right) $ & & $\phantom{=}$ & $\phantom{==} \left. + d_5 + d_7 + d_8 \right) $ \\ \hline
$\gamma_{B \pi K}^{(4)}$ & $8 \left(4d_1 - 2d_2 - d_5 \right.$  & $-16(d_7 - d_8)$ & $16\left( \frac{8}{3} d_3 + \frac{4}{3} d_4 - d_7 + d_8 \right) $ & $8 \left(4d_1 + 2d_2 + d_5 \right.$ \\ 
								&  $ \left.  - 2d_7 + 2d_8 \right)$ & & & $ \left.  - 2d_7 + 2d_8 \right)$  \\ \hline
\end{tabular}
}
\caption{Coefficients of the $\mathcal{O}(p^4)$ contact interactions.}	
\label{Table4}
\end{table}

The tadpole diagrams can be expressed as 
\begin{align}
  m_{B, \, tadpole}^{(4)} = \left[m_B^{(4)} \right]^{[1]} + \left[m_B^{(4)} \right]^{[2]}
  + \left[m_B^{(4)} \right]^{[3]} \; .
\end{align}
with
\begin{align}
\left[m_B^{(4)} \right]^{[1]} =  \frac{1}{(4 \pi F_{\phi})^2} \left\lbrace \xi_{B \pi}^{[1]} M_{\pi}^2 \log \left( \frac{M_{\pi}^2}{\mu^2} \right) + \xi_{B K}^{[1]} M_{K}^2 \log \left( \frac{M_{K}^2}{\mu^2} \right) \xi_{B \eta}^{[1]} M_{\eta}^2 \log \left( \frac{M_{\eta}^2}{\mu^2} \right) \right\rbrace \; ,
\end{align}
and
\begin{align}
\left[m_B^{(4)} \right]^{[2]} =  \frac{1}{(4 \pi F_{\phi})^2} \left\lbrace \xi_{B \pi}^{[2]} M_{\pi}^4 \log \left( \frac{M_{\pi}^2}{\mu^2} \right) + \xi_{B K}^{[2]} M_{K}^4 \log \left( \frac{M_{K}^2}{\mu^2} \right) + \xi_{B \eta}^{[2]} M_{\eta}^4 \log \left( \frac{M_{\eta}^2}{\mu^2} \right) \right\rbrace \; ,
\end{align}
and
\begin{align}
\begin{split}
\left[m_B^{(4)} \right]^{[3]} & = \frac{m_0}{(4 \pi F_{\phi})^2} \left\lbrace \xi_{B \pi}^{[3]} \left[ \frac{M_{\pi}^4}{4} \log \left( \frac{M_{\pi}^2}{\mu^2} \right) - \frac{M_{\pi}^4}{8} \right] + \xi_{B K}^{[3]} \left[ \frac{M_{K}^4}{4} \log \left( \frac{M_{K}^2}{\mu^2} \right) - \frac{M_{K}^4}{8} \right] \right. \\ 
&\phantom{= \frac{m_0}{(4 \pi F_{\phi})^2} =} \left. + \xi_{B \eta}^{[3]} \left[ \frac{M_{\eta}^4}{4} \log \left( \frac{M_{\eta}^2}{\mu^2} \right) - \frac{M_{\eta}^4}{8} \right] \right\rbrace \; ,
\end{split}
\end{align}
and the various coefficients are collected in Tab.~\ref{Table5}.
\begin{table}[h]
\centering
\resizebox{15cm}{!}{
\begin{tabular}{|c|c|c|c|c|} \hline
	& N & $\Sigma$ & $\Lambda$ & $\Xi$ \\ \hline
$\xi_{B \pi}^{[1]}$ & $3\left( 2b_0 + b_D + b_F \right) M_{\pi}^2 $ & $6 \left( b_0 + b_D \right) M_{\pi}^2 $ & $2 \left( 3b_0 + b_D \right) M_{\pi}^2 $ & $ 3\left( 2b_0 + b_D - b_F \right) M_{\pi}^2 $  \\
	& $\phantom{==}$ & $\phantom{=}  $ & $\phantom{=} $ & $\phantom{==} $ \\ \hline
$\xi_{B K}^{[1]}$ & $2 \left( 4b_0 + 3b_D - b_F \right) M_K^2 $ & $4 \left( 2b_0 + b_D \right) M_K^2 $ & $ \frac{4}{3}\left( 6b_0 + 5b_D \right) M_K^2$ & $2 \left( 4b_0 + 3b_D + b_F \right) M_K^2 $ \\
	&  $\phantom{==} $ & & $\phantom{=} $ & $\phantom{==}$ \\ \hline
$\xi_{B \eta}^{[1]}$ & $\frac{1}{3} \left[ 8(b_0 + b_D - b_F) M_K^2 \right.$  & $\frac{2}{3} \left[ 4 b_0 M_K^2 \right.$ & $\frac{2}{9} \left[ 4(3b_0 + 4b_D) M_K^2 \right.$ & $\frac{1}{3} \left[ 8(b_0 + b_D + b_F) M_K^2 \right. $ \\ 
	&  $ \left. - (2b_0 + 3b_D - 5b_F) M_{\pi}^2 \right]$ &  $ \left. + (b_D - b_F) M_{\pi}^2 \right] $ & $ \left. - (3b_0 + 7b_D) M_{\pi}^2 \right] $ & $\left. - (2b_0 + 3b_D + 5b_F) M_{\pi}^2 \right] $  \\ \hline \hline
$\xi_{B \pi}^{[2]}$ & $-3 \left( b_1 + b_2 + b_3 +2b_4 \right)$ & $-2 \left( 4b_1 + 2b_2 + 3b_4 \right)$ & $-2 \left( 2b_2 + 3b_4 \right)$ & $-3 \left( b_1 + b_2 - b_3 +2b_4 \right)$ \\
	& & & & \\ \hline
$\xi_{B K}^{[2]}$ & $-2 \left( 3b_1 + 3b_2 - b_3 +4b_4 \right)$ & $-4 \left( b_1 + b_2 + 2b_4 \right)$ & $-\frac{4}{3} \left( 9b_1 + b_2 + 6b_4 \right)$ & $-2 \left( 3b_1 + 3b_2 + b_3 +4b_4 \right)$ \\
	& & & & \\ \hline
$\xi_{B \eta}^{[2]}$ & $-\frac{1}{3} \left( 9b_1 + b_2 - 3b_3 +6b_4 \right)$ & $- \frac{2}{3} \left( 2b_2 + 3b_4  \right)$ & $-2 \left( 2b_2 + b_4 \right)$ & $-\frac{1}{3} \left( 9b_1 + b_2 + 3b_3 +6b_4 \right)$  \\ 
	& & & & \\ \hline \hline
$\xi_{B \pi}^{[3]}$ & $-6 \left( b_5 + b_6 + b_7 +2b_8 \right)$ & $-4 \left( 4b_5 + 2b_7 + 3b_8 \right)$ & $-4 \left( 2b_7 + 3b_8 \right)$ & $-6 \left( b_5 - b_6 + b_7 + 2b_8 \right)$ \\
& & & & \\ \hline
$\xi_{B K}^{[3]}$ & $-4 \left( 3b_5 - b_6 +  3b_7 +4b_8 \right)$ & $-8 \left( b_5 + b_7 + 2b_8 \right)$ & $-\frac{8}{3} \left( 9b_5 + b_7 + 6b_8 \right)$ & $-4 \left( 3b_5 + b_6 + 3b_7 +4b_8 \right)$ \\
& & & & \\ \hline
$\xi_{B \eta}^{[3]}$ & $-\frac{2}{3} \left( 9b_5 - 3b_6 + b_7 +6b_8 \right)$ & $- \frac{4}{3} \left( 2b_7 + 3b_8  \right)$ & $-4 \left( 2b_7 + b_8 \right)$ & $-\frac{2}{3} \left( 9b_5 + 3b_6 + b_7 + 6b_8 \right)$  \\ 
& & & & \\ \hline 
\end{tabular}
}
\caption{Coefficients of the $\mathcal{O}(p^4)$ tadpole diagrams.}	
\label{Table5}
\end{table}
Together with the contact interactions from Eq.~(\ref{ContactIntOp4}) there are $15$  LECs
from the meson-baryon Lagrangian entering the calculation.

The next contribution comes from the loop diagram with two baryon propagators.
First, we recall the HB result, see also  Refs.~\cite{Borasoy:1996bx,Frink:2004ic}.
The $\mathcal{O}(p^4)$ mass contribution in HB is given by
\begin{align}
\begin{split}
m_{B, \, HB \, loop}^{(4)} & = - \frac{1}{(4 \pi F_{\phi})^2} \left[ \alpha_{B \pi}^{(4)} M_{\pi}^2 \left( 2 + 3 \log \left( \frac{M_{\pi}^2}{\mu^2} \right) \right) + \alpha_{B K}^{(4)} M_{K}^2 \left( 2 + 3 \log \left( \frac{M_{K}^2}{\mu^2} \right) \right) \right. \\ 
& \phantom{= - \frac{1}{(4 \pi F_{\phi})^2} =} \left. + \alpha_{B \eta}^{(4)} M_{\eta}^2 \left( 2 + 3 \log \left( \frac{M_{\eta}^2}{\mu^2} \right) \right) \right] \; , 
\end{split}
\end{align}
with the coefficients from Tab.~\ref{Table6}. 
\begin{table}[t]
\centering
\resizebox{15cm}{!}{
\begin{tabular}{|c|c|c|c|c|} \hline
		& N & $\Sigma$ & $\Lambda$ & $\Xi$ \\ \hline
		$\alpha_{B \pi}^{(4)}$ & $-\frac{3}{4} (D+F)^2 m_{N}^{(2)}$ & $-\frac{1}{3} D^2 m_{\Lambda}^{(2)} - 2F^2 m_{\Sigma}^{(2)}$ & $-D^2 m_{\Sigma}^{(2)}$ & $-\frac{3}{4} (D-F)^2 m_{\Xi}^{(2)}$ \\ 
		& & & & \\ \hline
		$\alpha_{B K}^{(4)}$ & $- \frac{1}{12} (D + 3F)^2 m_{\Lambda}^{(2)}$ & $-\frac{1}{2} (D^2 + F^2) ( m_{N}^{(2)} + m_{\Xi}^{(2)} ) $ & $-\frac{1}{6} (D^2 + 9F^2) ( m_{N}^{(2)} + m_{\Xi}^{(2)} ) $ & $- \frac{1}{12} (D - 3F)^2 m_{\Lambda}^{(2)}$ \\ 
		& $- \frac{3}{4} (D-F)^2 m_{\Sigma}^{(2)}$ & $ + DF (m_{N}^{(2)} - m_{\Xi}^{(2)}) $ & $ - DF (m_{N}^{(2)} - m_{\Xi}^{(2)}) $ & $- \frac{3}{4} (D+F)^2 m_{\Sigma}^{(2)}$ \\ \hline
		$\alpha_{B \eta}^{(4)}$ & $-\frac{1}{12} (D - 3F)^2 m_{N}^{(2)}$ & $-\frac{1}{3} D^2 m_{\Sigma}^{(2)}$ & $-\frac{1}{3} D^2 m_{\Lambda}^{(2)}$ & $-\frac{1}{12} (D + 3F)^2 m_{\Xi}^{(2)}$ \\
		& & & & \\ \hline 
\end{tabular}
}
\caption{Coefficients of the $\mathcal{O}(p^4)$ self-energy diagram.}
\label{Table6}
\end{table}
Now we turn to the EOMS scheme. After some algebra, one can compactly express the fourth order
loop corrections  as
\begin{align}
  m_{B, \, loop}^{(4)} = \frac{1}{(4 \pi F_{\phi})^2} \sum_{\phi} \sum_{B'} \alpha_{B B' \phi}^{(4)}
  \mathcal{E}_{B' B}^{(4)} \left( M_{\phi} \right) \; ,
\end{align}
where $B' \in \lbrace N, \Sigma, \Lambda, \Xi \rbrace$, $\phi \in \lbrace \pi, K, \eta \rbrace$ and 
\begin{align}
\begin{split}
  \mathcal{E}_{B' B}^{(4)} \left( M_{\phi} \right) &= \frac{2M_{\phi}^3}{m_0^2 \sqrt{4m_0^2 - M_{\phi}^2}}
  \left[ 6m_0^2 \left( m_{B}^{(2)} - m_{B'}^{(2)} \right) - M_{\phi}^2 \left( 2m_{B}^{(2)} - m_{B'}^{(2)}
    \right) \right] \arccos \left( \frac{M_{\phi}}{2m_0} \right) \\
  &\phantom{=} \; - M_{\phi}^2 \left[ 2 \left( 2m_{B}^{(2)} - m_{B'}^{(2)} \right) + \left( m_{B}^{(2)}
    + m_{B'}^{(2)} \right) \log \left( \frac{m_0^2}{\mu^2} \right) \phantom{\frac{3m_0^2 \left( m_{B}^{(2)}
        - m_{B'}^{(2)} \right)  }{m_0^2}}  \right. \nonumber
\end{split}
\end{align}
\begin{align}
\begin{split}
    &\phantom{= = - M_{\phi}^2} \;  \left. + \frac{3m_0^2 \left( m_{B}^{(2)} - m_{B'}^{(2)} \right)
      - M_{\phi}^2 \left( 2m_{B}^{(2)} - m_{B'}^{(2)} \right) }{m_0^2} \log \left( \frac{M_{\phi}^2}{m_0^2}
    \right) \right] \; .
\end{split}
\end{align}
The coefficients $\alpha_{B B' \phi}^{(4)}$ are given in Tab.~\ref{Table7}.
\begin{table}[h]
\centering
\resizebox{15cm}{!}{
\begin{tabular}{|l|l|l|l|} \hline
N & $\Sigma$ & $\Lambda$ & $\Xi$ \\ \hline
$\alpha_{N N  \pi}^{(4)} = \frac{3}{4}(D+F)^2$ & $\alpha_{\Sigma \Sigma \pi}^{(4)} = 2F^2$ & & $\alpha_{\Xi \Xi  \pi}^{(4)} = \frac{3}{4}(D-F)^2$ \\
& & & \\ \hline
$\alpha_{N N  \eta}^{(4)} = \frac{1}{12}(D-3F)^2$ & $\alpha_{\Sigma \Sigma \eta}^{(4)} = \frac{1}{3}D^2$ & $\alpha_{\Lambda \Lambda \eta}^{(4)} = \frac{1}{3}D^2$ & $\alpha_{\Xi \Xi  \eta}^{(4)} = \frac{1}{12}(D+3F)^2$ \\
& & & \\ \hline
$\alpha_{N \Lambda K}^{(4)} = \frac{1}{12}(D+3F)^2$ & $\alpha_{\Sigma \Lambda \pi}^{(4)} = \frac{1}{3}D^2$ & $\alpha_{\Lambda \Sigma \pi}^{(4)} = D^2$ & $\alpha_{\Xi \Lambda K}^{(4)} = \frac{1}{12}(D-3F)^2$ \\
& & & \\ \hline
$\alpha_{N \Sigma K}^{(4)} = \frac{3}{4}(D-F)^2$ & $\alpha_{\Sigma N K}^{(4)} = \frac{1}{2}(D-F)^2$ & $\alpha_{\Lambda N K}^{(4)} = \frac{1}{6}(D+3F)^2$ & $\alpha_{\Xi \Sigma K}^{(4)} = \frac{3}{4}(D+F)^2$ \\ 
& & & \\ \hline
& $\alpha_{\Sigma \Xi K}^{(4)} = \frac{1}{2}(D+F)^2$ & $\alpha_{\Lambda \Xi K}^{(4)} = \frac{1}{6}(D-3F)^2$ & \\
& & & \\ \hline
\end{tabular}
}
\caption{Combined coefficients of the $\mathcal{O}(p^4)$ self-energy diagram.}
\label{Table7}
\end{table}

Finally, we must consider the decuplet contribution at this order, see the rightmost diagram
in the third row of Fig.~\ref{fig:diags}. In the HB approach, the fourth order baryon mass
shift takes the form
\begin{align}
\delta m_{B, \, HB}^{(4)} = \frac{\mathcal{C}^2}{64 \pi^2 F_{\phi}^2} \left\lbrace \beta_{B \pi}^{(4)} H_{\phantom{B}}^{(4)} (M_{\pi}) + \beta_{B K}^{(4)} H_{\phantom{B}}^{(4)} (M_{K}) + \beta_{B \eta}^{(4)} H_{\phantom{B}}^{(4)} (M_{\eta}) \right\rbrace \; ,
\end{align}
with 
\begin{align}
H_{\phantom{B}}^{(4)} (M_{\phi}) = \left( 4\Delta^2 - 2M_{\phi}^2 \right) \log \left( \frac{M_{\phi}}{\mu} \right) - M_{\phi}^2 -4 \Delta \sqrt{\Delta^2 - M_{\phi}^2} \log \left( \frac{\Delta}{M_{\phi}} + \sqrt{ \frac{\Delta^2}{M_{\phi}^2} -1 } \right) \; ,
\end{align}
for $M_{\phi}=M_{\pi}$ and 
\begin{align}
  H_{\phantom{B}}^{(4)} (M_{\phi}) = \left( 4\Delta^2 - 2M_{\phi}^2 \right) \log \left( \frac{M_{\phi}}{\mu}
  \right) - M_{\phi}^2 -4 \Delta \sqrt{M_{\phi}^2 - \Delta^2} \arccos \left( \frac{\Delta}{M_{\phi}} \right)\; ,
\end{align}
for $M_{\phi}=M_K, \, M_{\eta}$, and the  corresponding coefficients are given in Tab.~\ref{Table8}.
\begin{table}[h]
\centering
\begin{tabular}{|c|c|c|c|c|} \hline
		& N & $\Sigma$ & $\Lambda$ & $\Xi$ \\ \hline
		$\beta_{B \pi}^{(4)}$ & $8 m_{\Delta}^{(2)} $ & $\frac{4}{3} m_{\Sigma^*}^{(2)} $ & $6 m_{\Sigma^*}^{(2)} $ & $2 m_{\Xi^*}^{(2)}$ \\ 
		& & & & \\ \hline
		$\beta_{B K}^{(4)}$ & $2 m_{\Sigma^*}^{(2)}  $ & $\frac{4}{3} \left[ m_{\Xi^*}^{(2)} +4 m_{\Delta}^{(2)} \right] $ & $4 m_{\Xi^*}^{(2)} $ & $ 2 \left[ 2 m_{\Omega^-} + m_{\Sigma^*}^{(2)} \right] $ \\ 
		& & & & \\ \hline
		$\beta_{B \eta}^{(4)}$ & $0 $ & $ 2 m_{\Sigma^*}^{(2)}$ & $0 $ & $2 m_{\Xi^*}^{(2)}$ \\ 
		& & & & \\ \hline
\end{tabular}
\caption{Coefficients of the $\mathcal{O}(p^4)$ decuplet self-energy diagram.}
\label{Table8}
\end{table}
The covariant calculation is somewhat lengthy, we only give a short representation of the
final results. There are indeed two contributions, one form the genuine $\Order(p^4)$ diagram
just discussed in the HB approach and the other one from the leading covariant $\Order(p^3)$ diagram,
that also generates a fourth order correction. The first contribution takes the from
\begin{align}
\delta m_{B, 1}^{(4)} = \frac{\mathcal{C}^2}{13824 \pi^2 F_{\phi}^2} \left\lbrace \beta_{B \pi}^{(4)} \tilde{H}_{cov}^{(4), \, R} (M_{\pi}) + \beta_{B K}^{(4)} \tilde{H}_{cov}^{(4), \, R} (M_{K}) + \beta_{B \eta}^{(4)} \tilde{H}_{cov}^{(4), \, R} (M_{\eta}) \right\rbrace \; , \label{blub1}
\end{align}
with the coefficients from Tab.~\ref{Table6} and the function $\tilde{H}_{cov}^{(4), \, R} (M_{\phi})$ can be
found in Ref.~\cite{DSthesis}. The  $\mathcal{O}(p^4)$ contribution from the $\mathcal{O}(p^3)$ diagram
 is proportional to the second order octet-baryon masses. It reads
\begin{align}
\delta m_{B, 2}^{(4)} = \frac{m_B^{(2)} \mathcal{C}^2}{9216 \pi^2 F_{\phi}^2}  \left\lbrace \beta_{B \pi}^{(3)} \tilde{H}_{cov}^{(3,4), \, R} (M_{\pi}) + \beta_{B K}^{(3)} \tilde{H}_{cov}^{(3,4), \, R} (M_{K}) + \beta_{B \eta}^{(3)} \tilde{H}_{cov}^{(3,4), \, R} (M_{\eta}) \right\rbrace \; , \label{blub2}
\end{align}
with the coefficients from Tab.~\ref{Op3DecCoeff} and the function $\tilde{H}_{cov}^{(3,4), \, R} (M_{\phi})$
is also given in Ref.~\cite{DSthesis}.

We have thus completed the calculation of all the self-energies contributing to the octet-baryon mass
at chiral order $\mathcal{O}(p^4)$. Overall we have $22$ new LECs at fourth order, with $15$ coming
from the $\mathcal{O}(p^2)$ and $\mathcal{O}(p^4)$ meson-baryon Lagrangians, $5$ from the
$\mathcal{O}(p^4)$ meson Lagrangian and $2$ from the $\mathcal{O}(p^2)$ decuplet Lagrangian.

\subsubsection{$\sigma$-term}

Finally, we consider the sigma-term. It is calculated by using
\begin{align}
  \frac{\partial m_N}{\partial \hat{m}} = \left( \frac{\partial m_N}{\partial M_{\pi} } \right)
  \frac{\partial M_{\pi} }{\partial \hat{m}} + \left( \frac{\partial m_N}{\partial M_{K} } \right)
  \frac{\partial M_{K} }{\partial \hat{m}} + \left( \frac{\partial m_N}{\partial M_{\eta} } \right)
  \frac{\partial M_{\eta} }{\partial \hat{m}} \; .
\end{align}
This requires the Goldstone boson masses  at fourth order to calculate the derivatives of the
meson masses with respect to the quark masses $\hat{m}$ and $m_s$. Putting pieces together,
we obtain  for $\sigma_{\pi N}$ and $\sigma_{0}$ at fourth order 
\begin{align}
\begin{split}
\sigma_{\pi N}^{(4)} &= \frac{1}{2 M_{\pi}} \left\lbrace M_{\pi}^2 + \frac{16 M_{\pi}^4}{F_{\phi}^2} \left[ (2 L_6^r - L_4^r) + \frac{1}{2} (2 L_8^r - L_5^r) \right] +\frac{M_{\pi}^4}{36 \pi^2 F_{\phi}^2}  \right. \\
&\phantom{= \frac{1}{2 M_{\pi}} =} \left. + \frac{M_{\pi}^4}{32 \pi^2 F_{\phi}^2} \log \left(\frac{M_{\pi}^2}{\mu^2}\right) - \frac{M_{\pi}^4}{288 \pi ^2 F_{\phi}^2} \log \left(\frac{M_{\eta}^2}{\mu^2}\right) \right\rbrace \cdot \left(\frac{\partial m_N}{\partial M_{\pi}}\right) \\
&\phantom{=} + \frac{1}{2 M_{K}} \left\lbrace \frac{M_{\pi}^2}{2} + \frac{4 M_{\pi}^4}{F_{\phi}^2} \left[ 4(2 L_6^r - L_4^r) M_K^2 + (2 L_8^r - L_5^r) \left( 2 M_K^2 - M_{\pi}^2 \right)  \right] +\frac{M_K^2 M_{\pi}^2}{144 \pi^2 F_{\phi}^2} \right. \\
&\phantom{= + \frac{1}{2 M_{\pi}} =} \left. + \left[\frac{M_{\eta}^2 M_{\pi}^2}{64 \pi^2 F_{\phi}^2}+\frac{M_K^2 M_{\pi}^2}{144 \pi^2 F_{\phi}^2}\right] \log \left(\frac{M_{\eta}^2}{\mu^2}\right) - \frac{M_{\pi}^4}{64 \pi^2 F_{\phi}^2} \log \left(\frac{M_{\pi}^2}{\mu^2}\right) \right\rbrace \cdot \left(\frac{\partial m_N}{\partial M_{K}}\right) \\
&\phantom{=} + \frac{1}{2 M_{\eta}} \left\lbrace \frac{M_{\pi}^2}{288 \pi^2 F_{\phi}^2} \left[ 96 \pi^2 F_{\phi}^2+ 4608 \pi^2 (2 L_6^r - L_4^r) M_{\eta}^2-4096 \pi ^2 (3 L_7^r + L_8^r) M_{K}^2 \right. \right. \\ 
&\phantom{= + \frac{1}{2 M_{\pi}} =} \left. \left. +4096 \pi^2 (3 L_7^r + L_8^r) M_{\pi}^2+1536 \pi^2 (2 L_8^r - L_5^r) M_{\eta}^2-768 \pi ^2 (2 L_8^r - L_5^r) M_{\pi}^2 \right. \right. \\ 
&\phantom{= + \frac{1}{2 M_{\pi}} =} \left. \left. +5 M_{\eta}^2-5 M_{\pi}^2 \right] + M_{\pi}^2 \log \left(\frac{M_{\eta}^2}{\mu^2}\right)-4 M_{\eta}^2 \log \left(\frac{M_{\eta}^2}{\mu^2}\right) \right. \\ 
&\phantom{= + \frac{1}{2 M_{\pi}} =} \left. +3 \left(3 M_{\eta}^2+4 M_K^2+M_{\pi}^2\right) \log \left(\frac{M_{K}^2}{\mu^2}\right)-21 M_{\pi}^2 \log \left(\frac{M_{\pi}^2}{\mu^2}\right) \right\rbrace \cdot \left(\frac{\partial m_N}{\partial M_{\eta}}\right) \; , \label{eq854}
\end{split}
\end{align}
and
\begin{align}
\begin{split}
  \sigma_{0}^{(4)} &= \sigma_{\pi N}^{(4)} - \frac{1}{2 M_{\pi}} \left\lbrace \frac{M_{\pi}^4 \left(1152 \pi^2 (2 L_6^r - L_4^r) - \log \left(\frac{M_{\eta}^2}{\mu^2}\right)-1\right)}{72 \pi^2 F_{\phi}^2}  \right\rbrace \cdot \left(\frac{\partial m_N}{\partial M_{\pi}}\right) \\
&\phantom{=} - \frac{1}{2 M_{K}} \left\lbrace \frac{M_{\pi}^2}{288 \pi^2 F_{\phi}^2} \left[ 8 \left(36 \pi^2 \left(F_{\phi}^2-8 (2 L_8^r - L_5^r) M_{\pi}^2\right) \right. \phantom{\left(\frac{M_{\pi}^2}{\mu^2}\right)} \right. \right.
  \\
  &\phantom{= - \frac{1}{2 M_{K}} =} \left. \left. \left. + M_K^2 \left(576 \pi^2 (2 L_6^r - L_4^r) +576 \pi^2 (2 L_8^r - L_5^r)+1\right)\right) \phantom{\left(\frac{M_{\pi}^2}{\mu^2}\right)} \right. \right. \nonumber 
\end{split}
\end{align}
\begin{align}
\begin{split}
  &\phantom{= - \frac{1}{2 M_{K}} =} \left. \left. +\left(9 M_{\eta}^2+8 M_K^2\right) \log \left(\frac{M_{\eta}^2}{\mu^2}\right)-9 M_{\pi}^2 \log \left(\frac{M_{\pi}^2}{\mu^2}\right) \right] \right\rbrace \cdot \left(\frac{\partial m_N}{\partial M_{K}}\right) \\
&\phantom{=} - \frac{1}{2 M_{\eta}} \left\lbrace \frac{M_{\pi}^2}{144 \pi^2 F_{\phi}^2} \left[ 192 \pi^2 F_{\phi}^2+2304 \pi^2 (2 L_6^r - L_4^r) M_{\eta}^2+4096 \pi^2 (3 L_7^r + L_8^r) M_{K}^2  \phantom{\left(\frac{M_{\pi}^2}{\mu^2}\right)} \right. \right. \\
&\phantom{= - \frac{1}{2 M_{K}} =} \left. \left. -4096 \pi^2 (3 L_7^r + L_8^r) M_{\pi}^2+3072 \pi^2 (2 L_8^r - L_5^r) M_{\eta}^2 \phantom{\left(\frac{M_{\pi}^2}{\mu^2}\right)}   \right. \right. \\
  &\phantom{= - \frac{1}{2 M_{K}} =} \left. \left. -1536 \pi^2 (2 L_8^r - L_5^r) M_{\pi}^2 +5 M_{\pi}^2 +M_{\eta}^2 + 2 M_{\pi}^2 \log \left(\frac{M_{\eta}^2}{\mu^2}\right) \right. \right. \\ 
&\phantom{= - \frac{1}{2 M_{K}} =} \left. \left. -14 M_{\eta}^2 \log \left(\frac{M_{\eta}^2}{\mu^2}\right) +3 \left(3 M_{\eta}^2+4 M_K^2+M_{\pi}^2\right) \log \left(\frac{M_{K}^2}{\mu^2}\right) \right. \right. \\ 
&\phantom{= - \frac{1}{2 M_{K}} =} \left. \left. -6 M_{\pi}^2 \log \left(\frac{M_{\pi}^2}{\mu^2}\right) \right] \right\rbrace \cdot \left(\frac{\partial m_N}{\partial M_{\eta}}\right) \; . \label{eq855}
\end{split}
\end{align}
Using the explicit expressions for the fourth order nucleon mass in HB and covariant formulations (with and without the inclusion of the decuplet), we can calculate the derivatives of $m_N$ with respect to $M_{\pi}$, $M_K$ and $M_{\eta}$ to obtain the explicit expressions for the sigma-term and for $\sigma_{0}$.

\section{Fit procedure and error analysis}
\label{sec:fits}

We now have everything we need to calculate the quantity $\sigma_{0}$ from the nucleon mass up
to fourth chiral order. But our BCHPT results for the octet-baryon masses and the sigma-terms
contain LECs, whose values are unknown. So, in order to calculate $\sigma_{0}$ numerically and
deduce the strangeness content of the nucleon, we have to determine the LECs first. This is
achieved by fitting our BCHPT formulas to the physical values of the octet-baryon masses $N$,
$\Sigma$, $\Lambda$ and $\Xi$ and to the value of the sigma-term $\sigma_{\pi N}= (59.1 \pm 3.5)\,$MeV.
Clearly, these are not sufficient data to fix all LECs. Thus, we perform two types of fits, which
we call unconstrained and constrained fits, respectively, as will be discussed in what follows.

\subsection{Unconstrained fits}

We use the least-square-fitting procedure (often called $\chi^2$-fit) that is prominently used
in CHPT to determine the LECs for various processes, see e.g.
\cite{Bernard:1993nj,Yao:2016vbz,Siemens:2016hdi,Perez:2014yla}. 
The function $\chi^2(\lambda_1, \lambda_2, ..., \lambda_n)$ depends on $n$ parameters and is given by
\begin{align}
\chi^2(\lambda_1, \lambda_2, ..., \lambda_n) = \sum_{i=1}^{N}
\left[ \frac{O_i^{\text{exp}} - F_i(\lambda_1, \lambda_2, ..., \lambda_n)}{\Delta O_i^{\text{exp}}} \right]^2 \; ,
\label{eq91}
\end{align}
where $O_i^{\text{exp}}$ are $N$ experimental observables with their respective errors
$\Delta O_i^{\text{exp}}$. The expression $F_i(\lambda_1, \lambda_2, ..., \lambda_n)$ denotes the
theoretically evaluated function for the $i^{\rm th}$ observable, which depends on the unknown
parameters $\lambda_k$. The aim of the fit is to find the parameter values $\lambda_{k,0}$,
so that $\chi^2$ obtains a minimum 
\begin{align}
  \chi_{\text{min}}^2 = \min_{\lambda_k} \left\lbrace \chi^2(\lambda_1, ..., \lambda_n) \right\rbrace
  = \left. \chi^2(\lambda_1, ..., \lambda_n) \right\vert_{(\lambda_1, ..., \lambda_n) =
    (\lambda_{1,0}, ..., \lambda_{n,0}) } \; .
\end{align} 
Then the values $\lambda_{k,0}$ are our fit results.

We can also obtain the $n \times n$ error matrix $\mathbf{E}_{ij}$ for the fit by calculating the
inverse Hessian matrix of $\chi^2$ evaluated at the minimum, i.e.
\begin{align}
  \mathbf{E}_{ij}^{-1} = \frac{1}{2} \frac{\partial^2}{\partial \lambda_i \partial \lambda_j} \left.
  \left[ \chi^2 (\lambda_1, ..., \lambda_n) \right] \right\vert_{(\lambda_1, ..., \lambda_n)
    = (\lambda_{1,0}, ..., \lambda_{n,0})} \; .
\end{align}
The error of any parameter $\lambda_{k}$ is then given by the standard deviation
\begin{align}
\Delta \lambda_{k} = \sqrt{\mathbf{E}_{kk}} \; ,
\end{align}
and the correlation matrix is given by 
\begin{align}
\mathbf{C}_{ij} = \frac{\mathbf{E}_{ij}}{\sqrt{\mathbf{E}_{ii} \mathbf{E}_{jj}}} \; .
\end{align}
If we want to calculate another observable $G$, which also depends on the parameters $\lambda_{k}$,
we can use the error propagation formula
\begin{align}
  \left( \Delta G \right)^2 = \sum_{i,j} \left. \left( \frac{\partial G}{\partial \lambda_i} \right)
  \left( \frac{\partial G}{\partial \lambda_j} \right)\right\vert_{\lambda_k = \lambda_{k,0}} \cdot
  \mathbf{E}_{ij} \; . \label{errorpropagation}
\end{align}
In CHPT calculations, however, one is usually not only interested in errors from the fit, but also in
theoretical errors coming from the EFT approach itself. Since the results are obtained up
to a specific chiral order $\mathcal{O}(p^n)$, there are always corrections of $\mathcal{O}(p^{n+1})$,
that is not explicitly calculated. Thus an estimation is needed in order to see how much the result
changes if higher orders are included. The truncation uncertainty of an observable $G$, which
is calculated up to $\mathcal{O}(p^n)$, is given by \cite{Epelbaum:2014efa,Binder:2015mbz}
\begin{align}
  \left( \Delta G \right)_{\text{theo.}}^{(n)} = \max \left( | G^{(n_{\text{LO}})} | Q^{n-n_{\text{LO}}+1} ,
  \left\lbrace |G^{(k)} - G^{(j)}| Q^{n-j} \right\rbrace \right) \; ,
  \label{theoerror}
\end{align}
where $n_{\text{LO}}$ denotes the order of the leading-order result with $n_{\text{LO}} \le j < k \le n$
and $Q$ is the EFT expansion parameter given by 
\begin{align}
Q = \max \left( \frac{p}{\Lambda_{\chi}}, \frac{M_{\phi}}{\Lambda_{\chi}} \right) \; , \qquad
Q = \max \left( \frac{p}{\Lambda_{\chi}}, \frac{M_{\phi}}{\Lambda_{\chi}},
\frac{\Delta}{\Lambda_{\chi}} \right) \; , 
\end{align}
with the chiral symmetry breaking scale $\Lambda_{\chi} \simeq 1\,$GeV, where the first (second)
equation refers to the case without (with) the decuplet. In our analysis we calculate
the errors from the $\chi^2$-fit and the theoretical errors with $Q=M_{\eta}/\Lambda_{\chi} = 0.548$,
which is a rather conservative estimate.

Before we perform the fits, we have to remark a last detail about the $\mathcal{O}(p^4)$ fit. We have a
total of five experimental values ($m_N, m_{\Sigma}, m_{\Lambda}, m_{\Xi}, \sigma_{\pi N}$). At order
$\mathcal{O}(p^2)$ and $\mathcal{O}(p^3)$ there are four LECs ($m_0, b_0, b_D, b_F$), which can be fitted
very precisely, because we have more experimental values than LECs. However, at order $\mathcal{O}(p^4)$
there are $15$ new LECs from meson-baryon Lagrangians entering, but we still only have $5$ data to fit.
In order to find at least an estimation for the $\mathcal{O}(p^4)$ fit, we assume that the $15$ new
LECs have values around zero. We therefore use a so-called prior-fit and modify our
$\chi^2$ to, see e.g.~\cite{Schindler:2008fh},
\begin{align}
  \chi_{\text{prior}}^2 = \chi^2 + \sum_{i=1}^{8} \left(\frac{b_i}{\Delta b_i}\right)^2 +
  \sum_{i=1}^{5} \left(\frac{d_i}{\Delta d_i}\right)^2 + \left(\frac{d_7}{\Delta d_7}\right)^2
  + \left(\frac{d_8}{\Delta d_8}\right)^2 \; ,
  \label{eqpriorfit}
\end{align}
where $b_i$ are the LECs from the second order meson-baryon Lagrangian weighted by their
respective errors, $\Delta b_i$ and $d_i$ are the LECs from the fourth order Lagrangian with their
errors $\Delta d_i$. Adding these extra terms to the $\chi^2$, one ensures that the $15$ LECs obtain
values near zero (approximately Gaussian distributed), whereas the four LECs $m_0$, $b_0$, $b_D$
and $b_F$ are only fitted to the experimental data. Approximating the errors $\Delta b_i$ and
$\Delta d_i$ by a constant input value, we can minimize $\chi_{\text{prior}}^2$ to perform the
fit at $\mathcal{O}(p^4)$. 
Overall the prior-fit method behaves stable, since the minimum is not as flat as without the additional terms and all 19 LECs can be fitted quite precisely. However, we have to pay attention to choosing the errors $\Delta b_i$ and $\Delta d_i$. When the errors are too small the prior part of the fit will dominate over the actual $\chi^2$-fit, which is something we definitely intend to avoid.

\subsection{Constrained fits}

In the constrained fits, the $\chi^2$ function for the $\mathcal{O}(p^4)$ case is amended by additional terms
that reflect the constraints on some of the SU(3) LECs in terms of the well-known SU(2) LECs $c_{1,2,3}$
as discussed in Sec.~\ref{sec:constraints}. As before,
for the fits including the decuplet resonances, we have to subtract the $\Delta (1232)$-contribution from $c_2$ and $c_3$. 
We have to stress that including three new constraints from the SU(2) sector together with the octet-baryon masses and
$\sigma_{\pi N}$ is still not a sufficient amount of experimental values to fit all $19$ LECs with very good accuracy. 
This results in a fit, which is not as stable as the prior-fit method and does not allow a sensible error analysis. 
Therefore we will only present the central values of $\sigma_0$ obtained from this constrained fit. 

\section{Results and discussion}
\label{sec:results}

In our case the functions $F_i$ in Eq.~(\ref{eq91}) are our calculated octet-baryon masses and
the  result for $\sigma_{\pi N}$ from the previous sections. The unknown parameters $\lambda_k$
are the LECs from the second and fourth order meson-baryon Lagrangians that we want to determine in
order to calculate $\sigma_{0}$. We perform a fit for every order and for every case that we
calculated, i.e. an $\mathcal{O}(p^2)$, $\mathcal{O}(p^3)$ and $\mathcal{O}(p^4)$ fit for
the HB approach, the HB approach with the decuplet, the covariant calculation, and the
covariant calculation with the decuplet. We use the following data for the fits: 
The octet-baryon masses in the isospin limit\footnote{For a given multiplet, this is nothing but the sum of the masses divided by the number of states.} are $m_N = 938.9\,$MeV, $m_\Sigma = 1193.2\,$MeV,
$m_\Lambda = 1115.7\,$MeV and $m_\Xi = 1318.3\,$MeV. We use their respective standard
deviations as weighting factors for the $\chi^2$-fit. The standard deviation of the nucleon mass,
for example, is given by
\begin{align}
  \left( \Delta m_N \right) := \left[\frac{(m_{\text{proton}} - m_N)^2 + (m_{\text{neutron}}
      - m_N)^2}{2}\right]^{\frac{1}{2}} \; ,
\end{align}
with the PDG values for the proton and neutron mass. We obtain the standard deviations
$\Delta m_N = 0.65~{\rm MeV} \; ,  \Delta m_{\Sigma} = {3.30}~{\rm MeV} \; ,\Delta m_{\Xi} = {3.40}~{\rm MeV}$,
which will be used as our experimental uncertainties. Since there is only one $\Lambda$-baryon,
we assume $\Delta m_{\Lambda} = {1.00}~{\rm MeV}$. \\
As already mentioned, we use $\sigma_{\pi N}= (59.1 \pm {3.5})~{\rm MeV}$ for $\sigma_{\pi N}$ and its uncertainty.
The meson masses are  $M_{\pi}={139}~{\rm MeV}$ and $M_{K}={494}~{\rm MeV}$. Up to $\mathcal{O}(p^3)$ we
use the Gell-Mann--Okubo relation to obtain the $\eta$ mass, but at $\mathcal{O}(p^4)$ we use the PDG
value $M_{\eta}={548}~{\rm MeV}$. The other constants are given by
\begin{align}
F_{\phi} = 1.17 \cdot F_{\pi} \simeq 108~{\rm MeV} \; , \quad D=0.80  \; , \quad F=0.46 \; ,
\end{align}
$F_{\phi}$ the average of the pion, the kaon and the eta decay constants.
Inside the loop functions we set the average octet-baryon mass in the chiral limit $m_0$ equal
to $m_B={1.151}~{\rm GeV}$ (which is the average octet mass) and, analogously,
the decuplet mass to $m_D= 1.382~{\rm GeV}$ (which is the average decuplet mass).
The average
mass splitting between the octet and the decuplet, $\Delta$, is then given by 
$\Delta=m_D - m_B = {0.231}~{\rm GeV}$.
For the octet-decuplet-meson coupling $\mathcal{C}$, we use $\mathcal{C}=1.7$ 
and the off-shell parameter $z$ is set to $z=1/2$  in the covariant calculation. 
The LECs $L_4, L_5, L_6, L_7, L_8$ from the fourth order meson Lagrangian and $t_0, t_D$ from the
second order decuplet Lagrangian, which enter the baryon mass formulas at $\mathcal{O}(p^4)$,
have been obtained from other sources. The values for the former are from Ref.~\cite{Bijnens:2011tb},
while the results for the latter are given in App.~\ref{AppDecumasses}.
Throughout, we use  the renormalization scale $\mu=m_D={1.382}~{\rm GeV}$. 

The fit results for the LECs can be found in App.~\ref{AppFitresults} for  one example, namely the covariant
calculation without the decuplet at $\mathcal{O}(p^2)$, $\mathcal{O}(p^3)$ and $\mathcal{O}(p^4)$.
For the prior-fit we used the errors $\Delta b_{1,...,4}= {0.1}~{\rm GeV}^{-1}$,
$\Delta b_{5,...,8}= 0.1~{\rm GeV}^{-2}$ and $\Delta d_{1,...,5,7,8}= {0.1}~{\rm GeV}^{-3}$. We note that
the choice of larger uncertainties leads to a larger error.
With these values we calculate the quantity $\sigma_{0}$ and its errors using the error propagation
formula, Eq.~(\ref{errorpropagation}) and the theoretical truncation error estimation, Eq.~(\ref{theoerror}).
With the results from the $\mathcal{O}(p^2)$ fit, we obtain
\begin{align}
\sigma_{0} = 27.4 \left(0.2\right) \left(14.8\right)~{\rm MeV} \; ,
\end{align}
where the first bracket gives the uncertainty from the fit and the second gives the
theoretical uncertainty. The results for the higher chiral orders are given in Tab.~\ref{tab:res1}.
\begin{table}[t]
\centering
\resizebox{15cm}{!}{
\begin{tabular}{|c|c|c|c|c|}
\hline
               &    HB           &   HB + decuplet  &   EOMS         &   EOMS + decuplet \\ 
\hline
$\Order(p^3)$  & 57.9(0.2)(17.0) & 88.6(0.2)(34.0)  & 46.4(0.2)(10.4)& 57.6(0.2)(17.0) \\ 
$\Order(p^4)$  & 64.1(31.7)(9.3) & 64.0(31.7)(18.7) & 51.8(31.4)(5.7)& 61.8(31.4)(9.3) \\
\hline
\end{tabular}
}
\caption{Results for $\sigma_0$ in MeV at third and fourth order. Here, ``+decuplet''
  means the inclusion of the decuplet, HB denotes the heavy baryon and EOMS the covariant
  approach. The first error comes from the uncertainties within the given order, the
  second error is an estimate of the neglected higher order effects based on Eq.~(\ref{theoerror}).}
\label{tab:res1}
\end{table}
Similar results for the $\mathcal{O}(p^3)$ fit have been obtained earlier in \cite{Alarcon:2012nr}.  
The $\mathcal{O}(p^4)$ results for the covariant calculation only differ slightly from the
$\mathcal{O}(p^3)$ results. The fit without decuplet-resonances shifts more towards $\sigma_{\pi N}$,
while the fit with the decuplet is slightly above ${60}~{\rm MeV}$. The $\mathcal{O}(p^4)$ HB fit
results are both around ${64}~{\rm MeV}$. So the $\mathcal{O}(p^4)$ result for the HB approach
including decuplet baryons is much closer to $\sigma_{\pi N}$ than its corresponding $\mathcal{O}(p^3)$
result. This supports the idea that there are higher order terms contributing significantly,
which are not contained in the $\mathcal{O}(p^3)$ calculation.

Considering the errors of the calculation, we see that the errors from the fit are quite small
for the $\mathcal{O}(p^3)$ calculations. This is due to the fact that we have more experimental
data than LECs and thus the fit behaves well. The errors of the $\mathcal{O}(p^4)$ calculations
are very large because of the prior-fit method that we use. Since we assume that all $15$ LECs
from the $\mathcal{O}(p^4)$ calculation have the same error, we are not able to include any
correlations between them, which might reduce the error of $\sigma_{0}$. 

The theoretical error due to cutting off higher orders decreases overall from $\mathcal{O}(p^3)$
to $\mathcal{O}(p^4)$, as one would expect. The error from $\mathcal{O}(p^2)$, however, is
slightly smaller than most errors from $\mathcal{O}(p^3)$. The reason for this is that all
$\mathcal{O}(p^3)$ results (besides the covariant result without the decuplet) are close to
the $\sigma_{\pi N}$ value and differ by roughly ${30}~{\rm MeV}$ from the $\mathcal{O}(p^2)$
tree-level result. This implies that the  loop contribution plays an important role
in the calculation of $\sigma_{0}$. 

Given the central values for $\sigma_0$ from Tab.~\ref{tab:res1}, we see that the strangeness
content $y \simeq 0$, but due to the large uncertainties, this can not be made more precise. 

Consider now the constraint fit as described above. As already mentioned, we only give the central values for $\sigma_0$ here. For $c_{1,2,3}$ we use their numerical values given in Sec.~\ref{sec:constraints} and their respective errors as weighting factors for the fit. We also use $\bar{e}_1 = -1$~GeV$^{-3}$. Fits including the decuplet resonances are modified by subtracting the delta contribution from $c_2$ and $c_3$. Since the error of the $\Delta (1232)$-contribution is assumed to be quite large, see~\cite{Bernard:1996gq}, we adjust the errors of $(c_2-c_2^{\Delta})$ and $(c_3-c_3^{\Delta})$ to $\pm 1.0~{\rm GeV}^{-1}$. The constant $c_1$ remains unchanged. 

Overall we see that the value for $\sigma_0$ drops in comparison to the prior-fit results, cf. Tab.~\ref{tab:constr}. This also supports the idea of a vanishing strangeness content.

\begin{table}[h]
	\centering
	\begin{tabular}{|l|c|c|c|c|} \hline
	 & HB & HB + decuplet & EOMS & EOMS + decuplet \\ \hline
	 $\sigma_0$ $\left[ \text{MeV} \right]$ & $58.0$ & $60.4$ & $42.8$ & $60.9$ \\ \hline
	\end{tabular}
	\caption{Values of $\sigma_0$ for the $\mathcal{O}(p^4)$ constrained fits.}
	\label{tab:constr}
\end{table}

\section{Conclusions and outlook}\label{sec:conclusions}

In this paper, we have analyzed the ground-state octet baryon masses, the
pion-nucleon $\sigma$-term and its flavor decomposition to fourth order
in the chiral expansion, using the heavy baryon and the covariant EOMS
approach as well as including the contributions from the low-lying baryon decuplet.
We have entirely relied on continuum data and phenomenology to fix the
pertinent LECs. Here, we have mainly focused on the prediction for
$\sigma_0$, always using the value of $\sigma_{\pi N}= (59.1\pm 3.5)$~MeV from the
Roy-Steiner analysis of pion-nucleon scattering. The main findings of our investigation are: 
\begin{enumerate}
\item At third order, there is a large spread in the results for $\sigma_0$ depending
  on the BCHPT  scheme and whether or not the decuplet is included, cf. Tab.~\ref{tab:res1}.
  The uncertainties from the LECs within this order are very small. This confirms earlier
  findings of Ref.~\cite{Alarcon:2012nr}. In addition, we have shown that the error due
  to the neglect of higher orders is large.
\item  At fourth order, the central values of the fits are much closer, ranging from
  52~MeV to 64~MeV, consistent with a small strangeness fraction $y \simeq 0$.
  However, since not all LECs can be determined and the fit must be supplemented
  by a Bayesian ansatz for some of the LECs, the fit error within the order is
  sizeable, as expected. The uncertainty from neglecting higher orders is, however,
  much reduced compared to the third order.
\item Constraining certain combinations of the SU(3) LECs from matching to the
  SU(2) LECs, that are known to high precision from the Roy-Steiner analysis of pion-nucleon
  scattering, leads to a downward shift of a few MeV in the central value of $\sigma_0$.
\end{enumerate}

Once the apparent discrepancy in the determination of $\sigma_{\pi N}$ from Roy-Steiner
equations and the lattice is resolved, the formalism developed here can also be used to analyze
lattice results at varying quark masses. This should give a better handle on the badly
determined LECs. 

\section*{Acknowledgments}

We would like to thank Akaki Rusetsky for useful discussions.
Work supported in part by funds provided from the Deutsche Forschungsgemeinschaft
(SFB/TRR~110, ``Symmetries and the Emergence of Structure in QCD''), by the Chinese 
Academy of Sciences (CAS) President's International Fellowship Initiative (PIFI)
(grant no. 2018DM0034) and by VolkswagenStiftung (grant no. 93562).

\appendix

\section{Mesonic chiral Lagrangian}
\def\theequation{\Alph{section}.\arabic{equation}}
\setcounter{equation}{0}
\label{app:A}

For completeness, we discuss here the LO and NLO chiral Lagrangian of the Goldstone boson
fields. The LO Lagrangian is given by \cite{Gasser:1983yg,Gasser:1984gg},
\begin{align}
  \mathcal{L}_{\phi}^{(2)} = \frac{F_{\phi}^2}{4} \text{Tr} \left( D_{\mu} U ( D^{\mu} U )^{\dagger}
  \right) + \frac{F_{\phi}^2}{4} \text{Tr} \left( \chi U^{\dagger} + U \chi^{\dagger} \right) \; .
  \label{ML2}
\end{align}
Here, $F_{\phi}$ is the pseudoscalar decay constant in the chiral limit and the trace acts in
flavor space. The pseudoscalar fields are written in terms of a unitary $3 \times 3$ matrix $U$,
which is defined by 
\begin{align}
  U(x) = \exp \left( i \frac{\phi (x)}{F_{\phi}} \right) = \mathds{1} +
  i \frac{\phi (x)}{F_{\phi}} - \frac{\phi^2 (x)}{2 F_{\phi}^2} + ... \; ,
  \label{Umatrix}
\end{align}
with 
\begin{align}
	\phi (x) = \sum_{a=1}^{8} \lambda^a \phi^a(x) = \sqrt{2} \left(
	\begin{array}{ccc}
	\frac{1}{\sqrt{2}} \pi^0 + \frac{1}{\sqrt{6}} \eta & \pi^+ & K^+ \\
	\pi^- & - \frac{1}{\sqrt{2}} \pi^0 + \frac{1}{\sqrt{6}} \eta & K^0 \\
	K^- & \bar{K}^0 & - \frac{2}{\sqrt{6}} \eta \\
	\end{array}
	\right) \; .
\end{align}
The matrix $U$ transforms as $U \rightarrow R U L^{\dagger} $ under global $\text{SU}(3)_L \times
\text{SU}(3)_R$ transformations ($R, L \in \text{SU}(3)_{R,L}  $). $D_{\mu}$ is the covariant
derivative defined by 
\begin{align}
	D_{\mu} U := \partial_{\mu} U - i r_{\mu} U + i U l_{\mu} \; , 
\end{align}
with $r_{\mu} = v_{\mu} + a_{\mu}$ and $l_{\mu} = v_{\mu} - a_{\mu}$, where $v_{\mu}$ and
$a_{\mu}$ denote external vector and axial vector currents, respectively\footnote{We are
  only interested in strong interaction processes and therefore we set $v_{\mu} = a_{\mu} = 0$
  in the following.}. The covariant derivative transforms as $D_{\mu} U \to R (D_{\mu} U) L^{\dagger}$.
The second term in (\ref{ML2}) includes the explicit chiral symmetry breaking due to the
non-zero quark masses 
\begin{align}
	\chi = 2 B_0 \mathcal{M} \; , 
\end{align}
where $B_0$ is a constant related to the chiral quark condensate and $\mathcal{M} = \text{diag}
(m_u, m_d, m_s)$ is the quark mass matrix transforming as $\mathcal{M} \to R \mathcal{M} L^{\dagger}$.

At order $\mathcal{O}(p^4)$ (NLO) there are new terms that contribute to the meson Lagrangian 
\begin{align}
\begin{split}
  \mathcal{L}_{\phi}^{(4)} = \, & L_1 \left[ \text{Tr} \left( D_{\mu} U ( D^{\mu} U )^{\dagger} \right)
    \right]^2 + L_2 \text{Tr} \left( D_{\mu} U ( D_{\nu} U )^{\dagger} \right) \text{Tr} \left( D^{\mu}
  U ( D^{\nu} U )^{\dagger} \right) \\
  & + L_3 \text{Tr} \left( D_{\mu} U ( D^{\mu} U )^{\dagger} D_{\nu} U ( D^{\nu} U )^{\dagger} \right)
  + L_4  \text{Tr} \left( D_{\mu} U ( D^{\mu} U )^{\dagger} \right) \text{Tr} \left( \chi U^{\dagger}
  + U \chi^{\dagger} \right) \\
  & + L_5 \text{Tr} \left( D_{\mu} U ( D^{\mu} U )^{\dagger} ( \chi U^{\dagger} + U \chi^{\dagger} ) \right)
  + L_6 \left[ \text{Tr} \left( \chi U^{\dagger} + U \chi^{\dagger} \right) \right]^2 \\
  & + L_7 \left[ \text{Tr} \left( \chi U^{\dagger} - U \chi^{\dagger} \right) \right]^2
  + L_8 \text{Tr} \left( \chi U^{\dagger} \chi U^{\dagger} + U \chi^{\dagger} U \chi^{\dagger} \right)  + ... \; , \label{ML4}
\end{split}
\end{align}
where $L_1$, $L_2$, ..., $L_7$, and $L_8$ are LECs. The other terms of the $\mathcal{O}(p^4)$ Lagrangian,
that are not listed in Eq.~(\ref{ML4}) contain external vector and axial vector currents and thus are
not of relevance to our calculation.  There are also two terms without meson fields proportional
to the high-energy constants (HECs) $H_1$ and $H_2$. But they also do not contribute to our calculation.

\section{Decuplet-baryon masses at $\mathcal{O}(p^2)$}
\def\theequation{\Alph{section}.\arabic{equation}}
\setcounter{equation}{0}
\label{AppDecumasses}

The decuplet-baryon masses at $\mathcal{O}(p^2)$ can be calculated from the chiral symmetry breaking
Lagrangian in Eq.~(\ref{DecL2sb}). They are given by \cite{MartinCamalich:2010fp}
\begin{align*}
& m_{\Delta}^{(2)} = - (t_0 + 3 t_D) M_{\pi}^2 - 2 t_0 M_{K}^2 \; ,  \\
& m_{\Sigma^*}^{(2)} = - (t_0 + t_D) M_{\pi}^2 - (2 t_0 + 2 t_D) M_{K}^2  \; , \\
& m_{\Xi^*}^{(2)} = - (t_0 - t_D) M_{\pi}^2 - (2 t_0 + 4 t_D) M_{K}^2  \; , \\
& m_{\Omega^-}^{(2)} = - (t_0 - 3 t_D) M_{\pi}^2 - (2 t_0 + 6 t_D) M_{K}^2  \; .
\end{align*}
The LECs $t_0$ and $t_D$ can be determined from a fit to the $\mathcal{O}(p^3)$ decuplet-baryon masses.
Their values are $t_0={-0.27}~{\rm GeV}^{-1}$ and $t_D={-0.694}~{\rm GeV}^{-1}$ (assuming an average decuplet
mass of $m_D^{} = {1.382}~{\rm GeV}$ in the chiral limit)~\cite{MartinCamalich:2010fp}.

\section{Fit results}
\def\theequation{\Alph{section}.\arabic{equation}}
\setcounter{equation}{0}
\label{AppFitresults}

As one representative, we exhibit the values of the LECs for the covariant calculation without the
decuplet in Tab.~\ref{tab:LECscov}. The other fit results are given in Ref.~\cite{DSthesis}.

\begin{table}[h]
	\centering
	\begin{tabular}{|l|c|c|c|} \hline
		& Fit $\mathcal{O}(p^2)$ & Fit $\mathcal{O}(p^3)$ & Fit $\mathcal{O}(p^4)$ \\ \hline
		$m_0$ $\left[ \text{MeV} \right]$		 &$\left( 650.0 \pm 46.0 \right)$&$\left( 660.5 \pm 44.7 \right)$&$\left( 770.0 \pm 56.7 \right)$ \\ 
		$b_0$ $\left[ \text{GeV}^{-1} \right]$	&$\left( - 0.536 \pm 0.045 \right)$&$\left( - 0.873 \pm 0.045 \right)$&$\left( - 0.542 \pm 0.036 \right)$ \\ 
		$b_D$ $\left[ \text{GeV}^{-1} \right]$	&$\left(  0.063 \pm 0.003 \right)$&$\left(  0.063 \pm 0.003 \right)$&$\left( 0.081 \pm 0.020 \right)$ \\ 
		$b_F$ $\left[ \text{GeV}^{-1} \right]$	&$\left( - 0.216 \pm 0.001 \right)$&$\left( - 0.422 \pm 0.001 \right)$&$\left( - 0.313 \pm 0.003 \right)$ \\ 
		$b_1$ $\left[ \text{GeV}^{-1} \right]$	&		-			&		-		&$\left( -0.001 \pm 0.100 \right)$ \\ 
		$b_2$ $\left[ \text{GeV}^{-1} \right]$	&		-			&		-		&$\left( -0.001 \pm 0.100 \right)$ \\ 
		$b_3$ $\left[ \text{GeV}^{-1} \right]$	&		-			&		-		&$\left( -0.013 \pm 0.100 \right)$ \\ 
		$b_4$ $\left[ \text{GeV}^{-1} \right]$	&		-			&		-		&$\left( -0.003 \pm 0.100 \right)$ \\ 
		$b_5$ $\left[ \text{GeV}^{-2} \right]$	&		-			&		-		&$\left( 0.000 \pm 0.100 \right)$ \\ 
		$b_6$ $\left[ \text{GeV}^{-2} \right]$	&		-			&		-		&$\left( -0.007 \pm 0.100 \right)$ \\ 
		$b_7$ $\left[ \text{GeV}^{-2} \right]$	&		-			&		-		&$\left( 0.000 \pm 0.100 \right)$ \\ 
		$b_8$ $\left[ \text{GeV}^{-2} \right]$	&		-			&		-		&$\left( -0.001 \pm 0.100 \right)$ \\ 
		$d_1$ $\left[ \text{GeV}^{-3} \right]$	&		-			&		-		&$\left( 0.034 \pm 0.100 \right)$ \\ 
		$d_2$ $\left[ \text{GeV}^{-3} \right]$	&		-			&		-		&$\left( 0.016 \pm 0.100 \right)$ \\ 
		$d_3$ $\left[ \text{GeV}^{-3} \right]$	&		-			&		-		&$\left( 0.074 \pm 0.100 \right)$ \\ 
		$d_4$ $\left[ \text{GeV}^{-3} \right]$	&		-			&		-		&$\left( 0.055 \pm 0.100 \right)$ \\ 
		$d_5$ $\left[ \text{GeV}^{-3} \right]$	&		-			&		-		&$\left( 0.056 \pm 0.100 \right)$ \\ 
		$d_7$ $\left[ \text{GeV}^{-3} \right]$	&		-			&		-		&$\left( -0.096 \pm 0.100 \right)$ \\ 
		$d_8$ $\left[ \text{GeV}^{-3} \right]$	&		-			&		-		&$\left( 0.071 \pm 0.100 \right)$ \\ \hline
	\end{tabular}
	\caption{Fitted LECs are second, third and fourth order for the covariant calculation without the
                 decuplet as decribed in the text.}
        \label{tab:LECscov}
\end{table}



\end{document}